\documentclass[10pt,conference,compsocconf]{IEEEtran}

\usepackage{balance}
\usepackage{cite}
\usepackage{epsfig}
\usepackage{times}
\usepackage{graphicx}
\usepackage[figuresright]{rotating}
\usepackage{amssymb}
\usepackage{amsmath}
\usepackage{setspace}
\usepackage{rotating}
\usepackage{multirow}
\usepackage{wrapfig}
\usepackage{booktabs}
\usepackage{array}
\usepackage{comment}
\usepackage{pict2e}
\usepackage{epstopdf}
\usepackage{multirow}
\usepackage{hhline}
\usepackage{subcaption}
\usepackage[font=footnotesize]{caption}

\newcolumntype{L}[1]{>{\raggedright\let\newline\\\arraybackslash\hspace{0pt}}m{#1}}
\newcolumntype{C}[1]{>{\centering\let\newline\\\arraybackslash\hspace{0pt}}m{#1}}
\newcolumntype{R}[1]{>{\raggedleft\let\newline\\\arraybackslash\hspace{0pt}}m{#1}}

\usepackage{multirow,booktabs}
\usepackage[]{nohyperref}  %
\usepackage{url}  %

\usepackage[ruled]{algorithm2e}

\SetAlFnt{\small}
\SetAlCapFnt{\small}
\SetAlCapNameFnt{\small}
\SetAlCapHSkip{0pt}
\IncMargin{-\parindent}
\allowdisplaybreaks

\usepackage{times}
\usepackage{float}
\DeclareMathOperator*{\argmin}{\arg\!\min}

\usepackage{colortbl}

\newcommand{\be}{\begin{equation}}
\newcommand{\ee}{\end{equation}}

\begin{document}
\title{Analog Signal Compression and Multiplexing Techniques for Healthcare Internet of Things}
\author{\textbf{Xueyuan Zhao, Vidyasagar Sadhu, and Dario Pompili}\\
Department of Electrical and Computer Engineering, Rutgers University--New Brunswick, NJ, USA\\
E-mails: xueyuan\_zhao@cac.rutgers.edu, vidyasagar.sadhu@rutgers.edu, pompili@rutgers.edu}

\maketitle
\thispagestyle{empty}
\pagenumbering{gobble}

\begin{abstract}
Scalability is a major issue for Internet of Things~(IoT) as the total amount of traffic data collected and/or the number of sensors deployed grow. In some IoT applications such as healthcare, power consumption is also a key design factor for the IoT devices. In this paper, a multi-signal compression and encoding method based on Analog Joint Source Channel Coding~(AJSCC) is proposed that works fully in the analog domain without the need for power-hungry Analog-to-Digital Converters~(ADCs). Compression is achieved by quantizing all the input signals but one. While saving power, this method can also reduce the number of devices by combining one or more sensing functionalities into a single device (called `AJSCC device'). Apart from analog encoding, AJSCC devices communicate to an aggregator node (FPMM receiver) using a novel Frequency Position Modulation and Multiplexing~(FPMM) technique. Such joint modulation and multiplexing technique presents three mayor advantages---it is robust to interference at particular frequency bands, it protects against eavesdropping, and it consumes low power due to a very low Signal-to-Noise Ratio~(SNR) operating region at the receiver. 
Performance of the proposed multi-signal compression method and FPMM technique is evaluated via simulations in terms of Mean Square Error~(MSE) and Miss Detection Rate~(MDR), respectively.
\end{abstract}

\section{Introduction}
\textbf{Overview:} 
The novel paradigm of Internet of Things~(IoT) offers advanced connectivity of devices, systems and services, and will enable to go beyond current Machine-to-Machine~(M2M) communications while encompassing a variety of new protocols, domains, and applications. The interconnection of these embedded devices (including smart objects) is expected to cover a large amount of fields, while also enabling advanced applications like smart grids and expanding to futuristic domains such as smart cities and smart healthcare. The ``things'', in the IoT sense, can refer not only to a wide variety of sensing devices but also to objects and people; while devices in most cases only sense and are static, objects and people can also act on the environment and be mobile. Devices used in IoT range from heart monitoring implants, biosensors in body-networks, biochip transponders on farm animals, electric clams in coastal waters, to automobiles with built-in sensors to in-situ DNA sequencers for real-time, in-the-field environmental/food/pathogen monitoring. These ``things'' collect useful data with the help of existing technologies and then autonomously flow the data among the other devices.

\begin{figure}
\begin{center}
\includegraphics[width=3.5in]{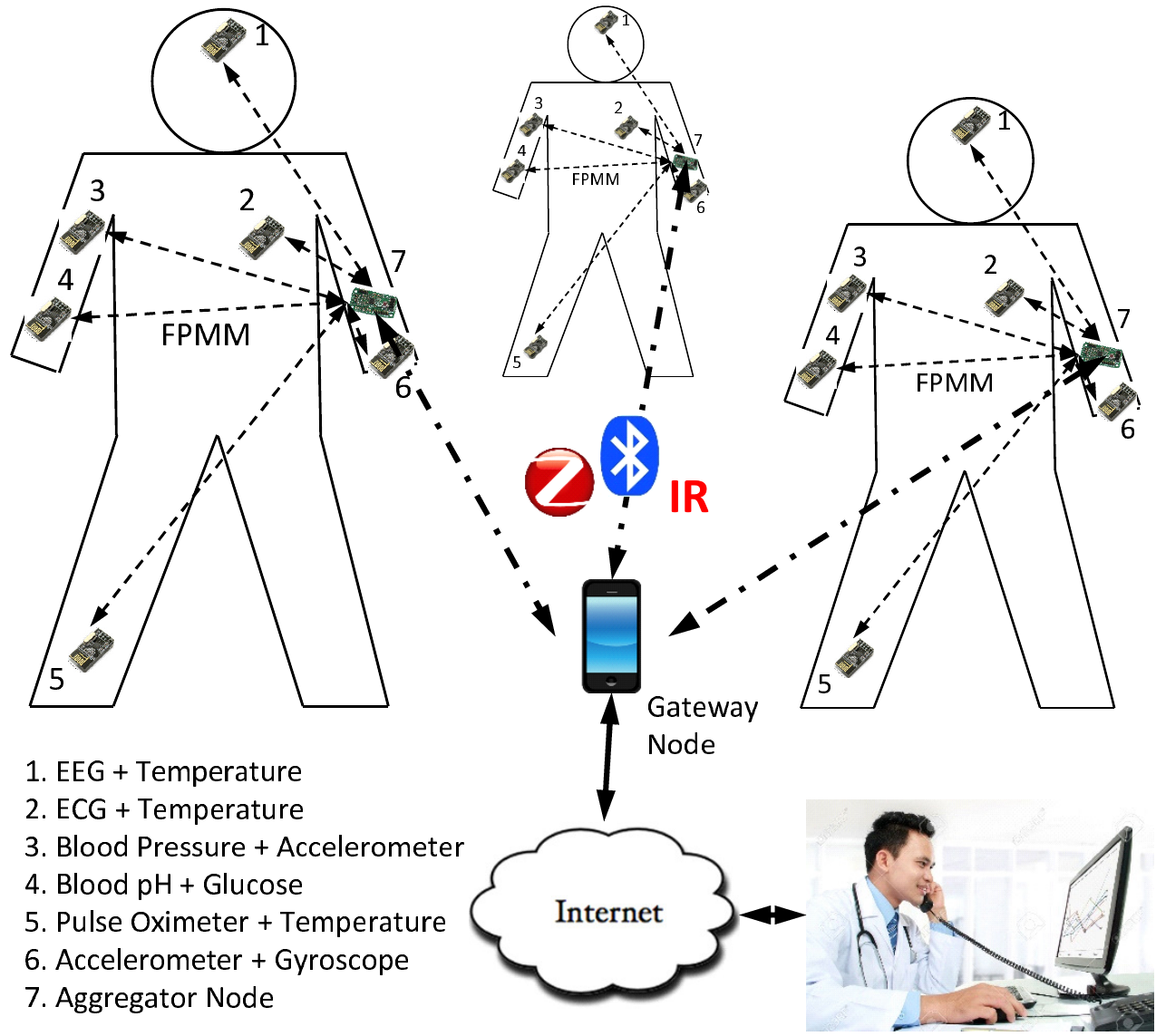}
\end{center}
\caption{AJSCC devices~(1-6) compress two or more sensor values into a single value and transmit to an Aggregator node~(7) using Frequency Position Modulation and Multiplexing~(FPMM). The Aggregator decodes individual values and transmits them to a Gateway (e.g., a mobile phone) using a Bluetooth/Zigbee/IR interface, which will then relay the data (individual/family) over Internet to the doctor. Upon analyzing the data, the doctor will increase/decrease the sensing granularity of the AJSCC IoT devices.}\label{fig:iot_healthcare}
\vspace{-0.2in}
\end{figure}

\textbf{Motivation:}
As the expansion of Internet-based automation has led to a plethora of new application domains and requirements, a denser and denser IoT is also expected to generate large amounts of data from diverse locations posing scalability as a major problem. In healthcare IoT, specifically, firstly people want to limit the number of body sensors worn so as to minimize discomfort. Secondly, with a large number of sensors transmitting simultaneously, better multiplexing techniques are needed for the sensors to communicate to a receiver node such as a Gateway that consumes less power and is robust to interference. Thirdly, with the miniaturization of sensors reaching new heights and the introduction of more resource-limited devices into the IoT arena, such as wearable/implantable sensors in Body Area Networks~(BAN), reducing power consumption of the sensing devices is a major challenge. Moreover, IoT for healthcare is important for the following reason: in the current clinical practice, a patient’s symptom may not appear when s/he visits the doctor, so the doctor often cannot objectively analyze it for diagnosis or treatment purposes. This leaves both the physician and the patient frustrated. If we could provide continuous monitoring with biomedical sensors that capture what is going on with the patient, we could understand why and under which conditions a symptom occurs. This example makes the case for healthcare IoT that continuously connects patients with their physician. 

\textbf{Our Approach:}
To address the data scalability problem, we propose a multi-signal \textit{compression} method that is based on Shannon mapping~\cite{Shannon49}, which is a low-complexity technique for Analog Joint Source-Channel Coding~(AJSCC)~\cite{Hekland05} that can compress two or more signals into one (introducing controlled distortion) while also staying resilient to wireless channel impairments. This technique also addresses the device scalability problem as two or more sensors can be integrated into a single device as the sensed values will be encoded/compressed using AJSCC. Furthermore, this technique will also address the power problem as encoding takes place in the analog domain without the need for power-hungry Analog-to-Digital~(ADC) converters (as we show below). To address the communication problem, we propose a novel Frequency Position Modulation and Multiplexing~(FPMM) technique to jointly modulate and multiplex the signals of a large number of sensors efficiently. We have designed it in such a way as to be robust to interference in particular frequency bands and to eavesdropping (more later). Due to low interference of the proposed FPMM technique, data from multiple devices is not mixed. Also, this technique can operate at a very low Signal-to-Noise Ratio~(SNR), thus greatly reducing the transmission power (data processing power is reduced by using analog encoding as mentioned earlier) compared to digital approaches. In summary, using our approach, body-worn devices can implement one or more sensing functionalities such as blood pH and glucose levels, or blood pressure and temperature, etc. The (heterogeneous) sensed signals are compressed/encoded in a single, compact device using AJSCC technique. The devices transmit the compressed values using FPMM to an aggregator node (FPMM receiver), which is assumed to have digital communication and processing capabilities. The aggregator decodes the data from the devices and transmits it to a Gateway node (e.g., a mobile phone) using traditional mobile short-range communication interfaces such as Bluetooth, Zigbee or IR, which can then connect to the Internet using Wi-Fi/LTE. 

\textbf{Broader Applications:}
In this paper we focus on healthcare IoT as an application of our approach; however, our approach is general and can be applied to other IoT scenarios as well. An example of IoT healthcare application using our approach is shown in Fig.~\ref{fig:iot_healthcare}, where different AJSCC devices~(1-6) transmit the sensor data to an Aggregator node~(7). The Aggregator decodes the data and transmits it to a Gateway node (e.g., a mobile phone), which is accessible to the doctor over the Internet. The doctor analyzes the patient's/entire family's data for group studies and can decide to increase or decrease the sensing granularity based on observed symptoms extracted from the data. For example, if no symptoms are detected, a lower granularity can be opted, thus saving resources.
We envision that our AJSCC devices' sensing granularity can also be programmed by the aggregator (i.e., they can receive some parameters too in addition to transmitting the sensed values). Hence, sensor data can be accessed \textit{anytime} and \textit{anywhere}, thus enabling innovative IoT applications where devices can not only sense but also receive configuration information from the cloud based on historical/statistical data. This configuration information can be obtained by mining the sensor data in the cloud. In this paper, however, we assume the sensors to be transmit only, leaving the more general case as future work.

\textbf{Our Contributions} are listed below: 
\begin{itemize}
\item We propose multi-signal compression techniques for IoT by extending the rectangular-type Shannon mapping to $N$ dimensions (N:1 compression). Closed-form expressions of the Mean Square Error~(MSE) have been derived.
\item We devise a Frequency Position Modulation and Multiplexing~(FPMM) technique to multiplex data from multiple sensors that is robust to interference at specific frequency bands, protects against eavesdropping, and has low power consumption.
\item We present numerical results that show the optimal parameters to minimize the MSE of the $N$-dimensional compression, and evaluate the proposed FPMM in terms of Miss Detection Rate~(MDR) by varying bandwidth, SNR, and number of multiplexed devices.
\end{itemize}

\textbf{Paper Outline:}
In Sect.~\ref{sec:rel_work}, we position our work w.r.t. others in healthcare IoT and analog sensing. In Sect.~\ref{sec:prop_soln}, we introduce and analyze the $N$-dimensional rectangular Shannon-mapping method (\ref{sec:signal-compress}) and explain our FPMM technique including how it reduces transmission power and how it provides robustness to eavesdropping and wireless channel interference (\ref{sec:fpm}). In Sect.~\ref{sec:perf_eval}, we present performance simulation results of MSE for the $N$-dimensional compression method and of MDR for the proposed FPMM technique. Finally, in Sect.~\ref{sec:conc}, we draw our conclusions and indicate future directions.

\begin{table}[t]
  \centering
  \caption{Power comparison with state-of-the-art wireless sensor motes.}\label{table:compare}
  \footnotesize
\begin{tabular}{|m{4cm}|m{3.5cm}|}
\hline
\textbf{Wireless sensor}                                                             & \textbf{Comparison with ours}                                                                           \\ \hline \hline
WSN340~\cite{wsn340}: Active MCU power consumption of 1.1~$\mathrm{mW}$        & \multirow{4}{*}{\scriptsize \parbox{3.5cm}{Our sensor~\cite{Zhao16} will consume $\approx130~\mu{\mathrm{W}}$ with state-of-the-art low-power components (OpAmps, etc.). There is potential to reduce this even further ($<50~{\mu \mathrm{W}}$) when all the functionalities are integrated into a monolithic component using analog IC design.}} \\ \cline{1-1}
Mantaro CoSeN~\cite{cosen}: Active MCU power consumption of 2.4~$\mathrm{mW}$ &                                                                             \\ \cline{1-1}
Telos RevB~\cite{telosb}: Active MCU power consumption of 6~$\mathrm{mW}$     &                                                                            \\ \cline{1-1}
MICA2~\cite{mica2}: Active MCU power consumption of 26.4~$\mathrm{mW}$        &                                                                            \\ \hline
\end{tabular}
  \vspace{-0.2in}
\end{table}

\section{Related Work}\label{sec:rel_work}
In the IoT healthcare and BAN domain, all of the existing solutions do sensing and communication in the digital domain (using ADCs/DACs/Microprocessors), which needs more power than analog sensing and communication. For example, Yang's group~\cite{2014BodyNetworks} developed a miniaturized node that incorporates wireless communication, on-board processing, nine-axis motion tracking, and other sensors~\cite{Andreu-Perez2015FromHealthcare}. It also developed an e-AR sensor~\cite{Jarchi2015AssessmentPatients}, a small device to be worn behind the ear that captures information about the balance of the wearer such as gait, posture, skeletal/joint shock-wave transmission, and activity of the individual. Yuce's group developed techniques based on Ultra Wide Band~(UWB) wireless technology to reduce the power consumption of body-worn sensors~\cite{Thotahewa2014PowerPath, Yuce2015WE-Harvest:Harvester}. 
Another important example is the activity recognition of the user using various body sensors viz., accelerometer and gyroscope data. The data from the sensors is digitally processed and stored into a wrist-band device that syncs the data with the mobile phone using Bluetooth technology~\cite{Chetty2016BodyRecognition}. 
Unlike these approaches, we adopt an entirely different approach based on analog sensing and communication that does not use any power-hungry ADCs (see Table~\ref{table:compare}).

Shannon mapping has been applied in a number of applications such as Software-Defined Radio~(SDR) systems~\cite{Garcia11}, optical digital communications~\cite{Romero14}, Compressive Sensing~(CS)~\cite{Saleh12}, and digital video transmissions~\cite{stopler14}. All these applications use power-hungry ADCs and other digital components making such implementations unsuitable for healthcare and other low-power IoT solutions. Some works have studied the N:1 spiral-type mapping~\cite{Brante13}. 
The advantage of considering rectangular-type Shannon mapping is that there are existing low-power, all-analog circuit realizations for rectangular-type mapping (our previous work~\cite{Zhao16}). Using this approach, sensors can be designed using all-analog circuits that can compress multiple signals into one signal, thereby consuming less power. The signals from multiple sensors are multiplexed at different frequency locations in an interleaved pattern. Similar pattern has been studied for topics of pilot placement~\cite{Zhao06, Zhao07}. 

This paper is the first work to propose this structure for multiplexing data of AJSCC sensors. Table~\ref{table:compare} compares the power numbers of our circuit~\cite{Zhao16} with state-of-the-art \textit{low-power} wireless sensor motes (all of which are digital). We can notice that $<50\mu W$ is possible with our circuit, which is essential in low-power applications. The existing circuit realizations of spiral-type mapping also are all based on digital circuits and systems~\cite{Garcia11}. In this scenario, it is worth noting that the Hybrid Digital Analog~(HDA) coding can also perform signal compression~\cite{Abbasi14,Behroozi11}. However, the digital part still needs digitization of the signals. Contrary to all these approaches, we propose signal compression and encoding in the analog domain with no need for ADCs/DACs/Microprocessors. To show the feasibility of our vision (analog compression and encoding), we previously developed a novel \textit{analog} circuit to compress two signals~\cite{Zhao16} and verified its applicability to two pathological signals (molecular biomarkers and physiological signal)~\cite{Zhao17}. In this paper, we extend the theory for N-dimensional signal compression and propose novel multiplexing techniques that address the above mentioned challenges of scalability and power in the context of healthcare IoT as one of the key applications.

\section{Proposed Solution}\label{sec:prop_soln}
In this section, we present our novel techniques for N-dimensional signal compression first (Sect.~\ref{sec:signal-compress}) and then for joint signal modulation and multiplexing (Sect.~\ref{sec:fpm}).

\subsection{N-dimensional Signal Compression}\label{sec:signal-compress}
We introduce and present the motivation for using Analog Joint Source Channel Coding~(AJSCC); then, 
we derive the theory to compute the MSE for both the 3- and N-dimensional rectangular-type Shannon mapping.
Note that the mathematical analysis on high-dimension rectangular-type Shannon mapping is not trivial and has not been studied before.

\subsubsection{Analog Joint Source Channel Coding~(AJSCC)}\label{sec:ajscc}
AJSCC adopts Shannon mapping as its encoding method~\cite{Hekland05}. Such mapping, in which the design of \emph{rectangular (parallel) lines} can be used for 2:1 compression, was first introduced by Shannon in his seminal 1949 paper~\cite{Shannon49}. Later work has extended this mapping to a \emph{spiral type} as well as to N:1 mapping~\cite{Brante13}. AJSCC requires simple compression and coding, and low-complexity decoding. 

To compress the source signals (``sensing source point"), the point on the space-filling curve with minimum Euclidean distance from the source point is found (``AJSCC mapped point"). The two most-widely adopted mapping methods are rectangular and spiral shaped: in the former, the transmitted signal is the ``accumulated length'' of the lines from the origin to the mapped point; while in the latter it is the ``angle'' that \emph{uniquely} identifies the mapped point on the spiral. At the receiver---a Cluster Head~(CH)---the reverse mapping is performed on the received signal using Maximum Likelihood~(ML) decoding. The error introduced by the two mappings is controlled by the spacing $\Delta_H$ between lines and spacing $\Delta_S$ between spiral arms, respectively: with smaller $\Delta_H$ (or $\Delta_S$), approximation noise is reduced; however, channel noise is increased as a little variation can push the received symbol to the next parallel line (or spiral arm). In addition, the mapping signal range would also increase, pushing more resources for transmission. Hence, Shannon mapping has the two-fold property of (1)~compressing the sources (by means of N:1 mapping) and (2)~being robust to (wireless) channel distortions as the noise only introduces errors along the parallel lines (or the spiral curve). Note that AJSCC performs analog compression at the symbol level; the fact that symbols are memoryless makes it a low-latency and low-complexity solution that is very suitable for practical implementations. For more supporting information, refer to our work in~\cite{wons3tier2017}.

\begin{figure}[t]
\begin{center}
\includegraphics[width=3.7in]{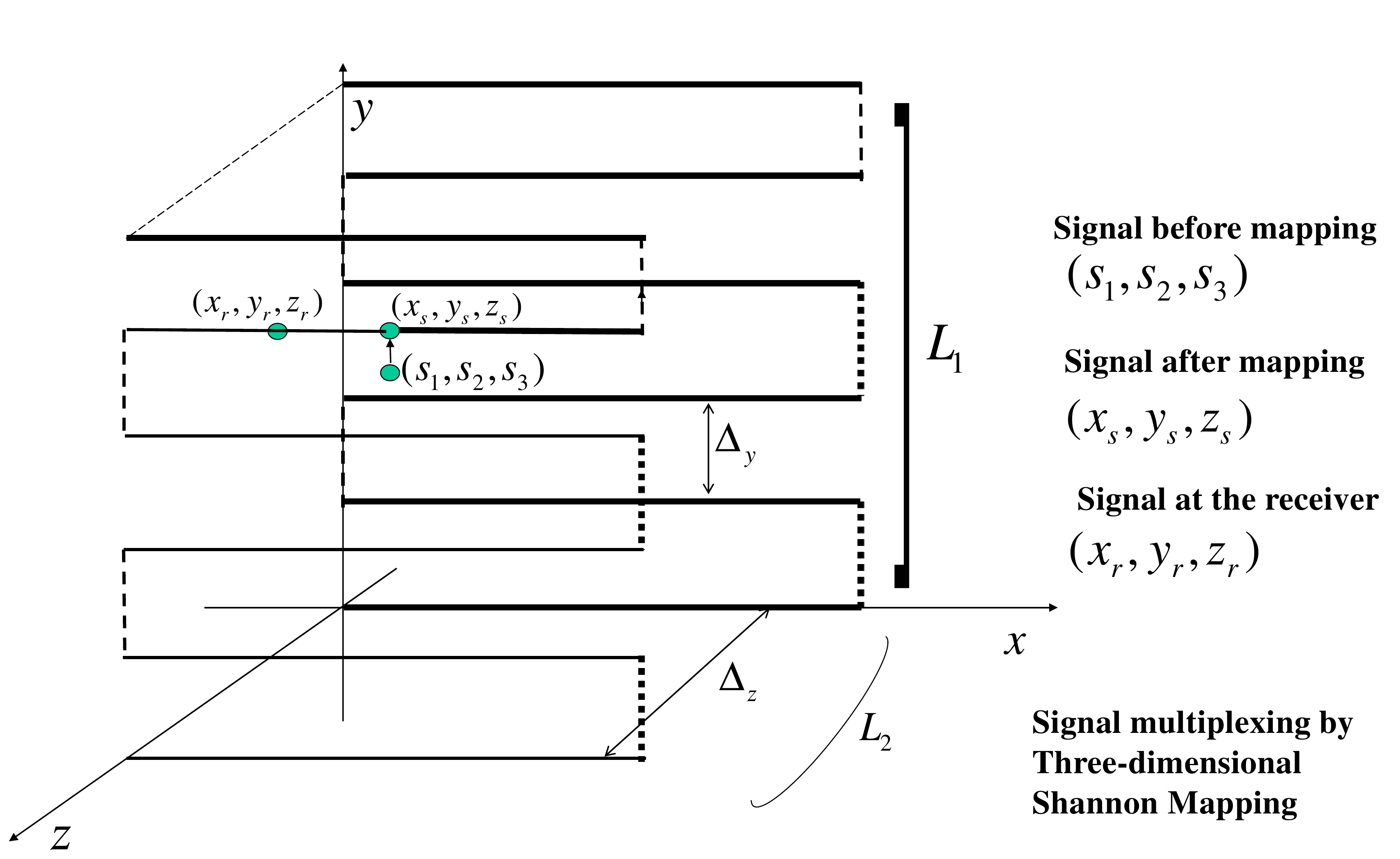}
\end{center}
\vspace{-1.0\baselineskip}
\caption{Signal compression in 3:1 Shannon mapping.}\label{fig:signal_multiplexing}
\vspace{-0.2in}
\end{figure}

\subsubsection{Analysis of N:1 Mapping}\label{sec:high_snr}
For simplicity, we will first start by analyzing the 3:1 mapping; then, we will generalize it to N:1.
The 3:1 mapping of rectangular type is depicted in Fig.~\ref{fig:signal_multiplexing}. The mapped signal is the accumulated length of the lines from the origin to the mapped point.
Note that, since the output is an analog electric signal (such as voltage), there will be an upper limit on this signal's amplitude (such as the circuit's supply voltage), which will limit the number of signals that can be simultaneously multiplexed.

Let us define the source analog signal as $(s_1, s_2, s_3)$, which is a continuous signal with a distribution $p_{S1}$, $p_{S2}$, and $p_{S3}$. The distributions are bounded with range [$Q_{1,L}$,$Q_{1,R}$], [$Q_{2,L}$,$Q_{2,R}$], and [$Q_{3,L}$,$Q_{3,R}$]. At the transmitter, the source signals are firstly mapped to $(a_1, a_2, a_3)$. The operation on $s_1$ is to shift the distribution to the positive value range [0,$R_1$], where $R_1 = Q_{1,R} - Q_{1,L}$, then linearly scale the distribution to the range of $[0,d]$, where $d$ is the length of each line. The operation on $s_2$ and $s_3$ are to shift the distribution to the range [0,$R_2$] and [0,$R_3$], respectively, where $R_2$ = $Q_{2,R}$-$Q_{2,L}$ and $R_3$ = $Q_{3,R}$-$Q_{3,L}$. The second step at the transmitter is to map the signal to one point on the parallel lines so as to produce the point of $(x_s, y_s, z_s)$. In this mapping, there is quantization error introduced to $y_s$ and $z_s$. The value of $x_s$ is the same of $a_1$. At the receiver, due to channel noise, the received signal will be perturbed from its original point.

Assuming medium to high SNR, the probability that the received point moves to another parallel line is so small that it can be neglected in the analysis. For Gaussian noise and in such SNR regime, the received signal can be expressed by $x_r = x_s + n$, $y_r = y_s$, and $z_r = z_s$. Assume the transmitted signal has the maximum amplitude $D_{max}$, then $d = D_{max}/(L_1 L_2)$. The recovered source signal is denoted by $(\tilde s_1, \tilde s_2, \tilde s_3)$, where $\tilde s_1$ can be written as,
\begin{equation}\label{eq:s_1}
 \tilde s_1  = \frac{{R_1 }}{d}x_r  = \frac{{R_1 L_1 L_2 }}{{D_{\max } }}(x_s  + n)= s_1  + \frac{{R_1 L_1 L_2 }}{{D_{\max } }}n.
 \end{equation}From~\eqref{eq:s_1} we see that the greater $L_1$ and $L_2$, the larger the error. The received signal $\tilde s_2$ can be written as $\tilde s_2  = y_s  = s_2  + \lambda _y$, where $\lambda _y$ is the error term. Based on the mapping method, $y_s  = round\left( {\frac{{s_2 }}{{\Delta _y }}} \right) \cdot \Delta _y$, where $\Delta _y$ is the spacing of the lines in the y-axis and has the value of $R_2 / (L_1 - 1)$. Hence, the error term $\left| {\lambda _y } \right|$ can be written as,
\begin{equation}
\left| {\lambda _y } \right| = \left| {s_2  - \tilde s_2 } \right| = \left| {s_2  - round\left( {\frac{{s_2 }}{{\Delta _y }}} \right) \cdot \Delta _y } \right|.
\end{equation}
If we assume now a uniform source signal distribution of $s_2$, the $\left| {\lambda _y } \right|$ is a random variable (r.v.) uniformly distributed in $[0, {\Delta _y }/2]$. Similarly, we have $\tilde s_3  = z_s  = s_3  + \lambda _z$, where the error term $\left| {\lambda _z } \right|$ is,
\begin{equation}
\left| {\lambda _z } \right| = \left| {s_3  - \tilde s_3 } \right| = \left| {s_3  - round\left( {\frac{{s_3 }}{{\Delta _z }}} \right) \cdot \Delta _z } \right|,
\end{equation}and $\Delta _z$=$R_3 / (L_2 - 1)$. The sum MSE of the three signals is,
\begin{equation}
 MSE = E\{ {\left| {s_1  - \tilde s_1 } \right|_{}^2 } \} + E\{ {\left| {s_2  - \tilde s_2 } \right|_{}^2 } \} + E\{ {\left| {s_3  - \tilde s_3 } \right|_{}^2 } \},
\end{equation}which can be further written as,
\begin{equation}\label{eq:MSE}
MSE = \frac{{R_1^2 L_1^2 L_2^2 }}{{D_{\max }^2 }}E\{ \left| n \right|_{}^2 \}  + E\{ {\left| {\lambda _y } \right|_{}^2 } \} + E\{ {\left| {\lambda _z } \right|_{}^2 } \}.
\end{equation}
The Probability Density Function~(PDF) of $\left| {\lambda _y } \right|$ is,
\begin{equation}
p_{\left| {\lambda_y } \right|}  = \left\{ \begin{array}{l}
 2/\Delta _y ,{\kern 1pt} {\kern 1pt} {\kern 1pt} {\kern 1pt} {\kern 1pt} {\kern 1pt} {\kern 1pt} {\kern 1pt} {\kern 1pt} {\kern 1pt} {\kern 1pt} \left| {\lambda _y } \right| \in [0,\Delta _y /2]{\kern 1pt} {\kern 1pt}  \\
 0{\kern 1pt} {\kern 1pt} {\kern 1pt} {\kern 1pt} {\kern 1pt} {\kern 1pt} {\kern 1pt} {\kern 1pt} {\kern 1pt} {\kern 1pt} {\kern 1pt} {\kern 1pt} {\kern 1pt} {\kern 1pt} {\kern 1pt} {\kern 1pt} {\kern 1pt} {\kern 1pt} {\kern 1pt} {\kern 1pt} {\kern 1pt} {\kern 1pt} {\kern 1pt} ,{\kern 1pt} {\kern 1pt} {\kern 1pt} {\kern 1pt} {\kern 1pt} {\kern 1pt} {\kern 1pt} {\kern 1pt} {\kern 1pt} {\kern 1pt} {\kern 1pt} otherwise.{\kern 1pt} {\kern 1pt} {\kern 1pt} {\kern 1pt}  \\
 \end{array} \right.
\end{equation}
The expectation of $\left| {\lambda _y } \right| ^2$ in~\eqref{eq:MSE} is thus written as,
\begin{equation}
E\{ {\kern 1pt} \left| {\lambda _y } \right|_{}^2 \}  = \int\limits_{}^{} {{\kern 1pt} \left| {\lambda _y } \right|_{}^2 } p_{\left| {\lambda _y } \right|} d{\left| {\lambda _y } \right|}= \frac{1}{{12}}\Delta _y^2  = \frac{1}{{12}}\frac{{R_2^2 }}{{(L_1^{}  - 1)_{}^2 }}.
\end{equation}
Similarly, the expectation of $\left| {\lambda _z } \right| ^2$ in~\eqref{eq:MSE} is,
\begin{equation}
E\{ {\kern 1pt} \left| {\lambda _z } \right|_{}^2 \}  = \frac{1}{{12}}\frac{{R_3^2 }}{{(L_2^{}  - 1)_{}^2 }}.
\end{equation}Since noise $n$ follows a Normal distribution $\mathcal{N}(0,\sigma_n^2)$, we have $E\{ \left| n \right|_{}^2 \}=\sigma _n^2$. Finally, the MSE for 3:1 mapping can be expressed in the following closed form, i.e.,
\begin{equation}
 MSE = \frac{{R_1^2 L_1^2 L_2^2 }}{{D_{\max }^2 }}\sigma _n^2  + \frac{1}{{12}}\frac{{R_2^2 }}{{(L_1^{}  - 1)_{}^2 }}+\frac{1}{{12}}\frac{{R_3^2 }}{{(L_2^{}  - 1)_{}^2 }}.
\end{equation}

We can now extend this result to the N:1 mapping. 
The source signal $(s_1 ,s_2 ,...,s_N )$ having range $[0,R_1]$, $[0,R_2]$, ..., $[0,R_N]$ is scaled and mapped to a point, $(x_1, x_2, ..., x_N)$, on the $N$-dimensional space by $x_1  = \frac{d}{{R_1 }}s_1$, $x_2  = s_2  + \lambda _2$, ..., $x_N  = s_N  + \lambda _N {\kern 1pt}$. The transmitting signal constraint now generalizes to $\mathop \prod \limits_{k = 1}^{N - 1} L_k^{}  \cdot d \le D_{\max}$. Similarly to the 3:1 mapping analysis, the MSE for N:1 mapping can be expressed as,
\begin{equation}
MSE_N = \left( {\frac{{R_1^{} \mathop \prod \limits_{k = 1}^{N - 1} L_k^{} }}{{D_{\max }^{} }}} \right)_{}^2 \sigma _n^2  + \sum\limits_{k = 1}^{N - 1} {\frac{1}{{12}}\frac{{R_{k + 1}^2 }}{{(L_k^{}  - 1)_{}^2 }}}.
\end{equation}

The MSE minimization problem, which can be solved numerically, can be written as,
\begin{equation}\label{eq:opt}
\mathop{\argmin}\limits_{L_1^{} ,L_{2,...}^{} ,L_{N - 1}^{} } MSE_N,
\end{equation}where in~\eqref{eq:opt} the number of stages per dimension needs to be an integer greater than 1, i.e., $L_k\in \mathbb{N}\setminus\{1\}, \forall k=1...N-1$.

\begin{figure*}[t!]
        \centering   
           \begin{subfigure}[b]{0.30\textwidth}
        		\centering
        		\includegraphics[width=1\textwidth, height=3.5cm]{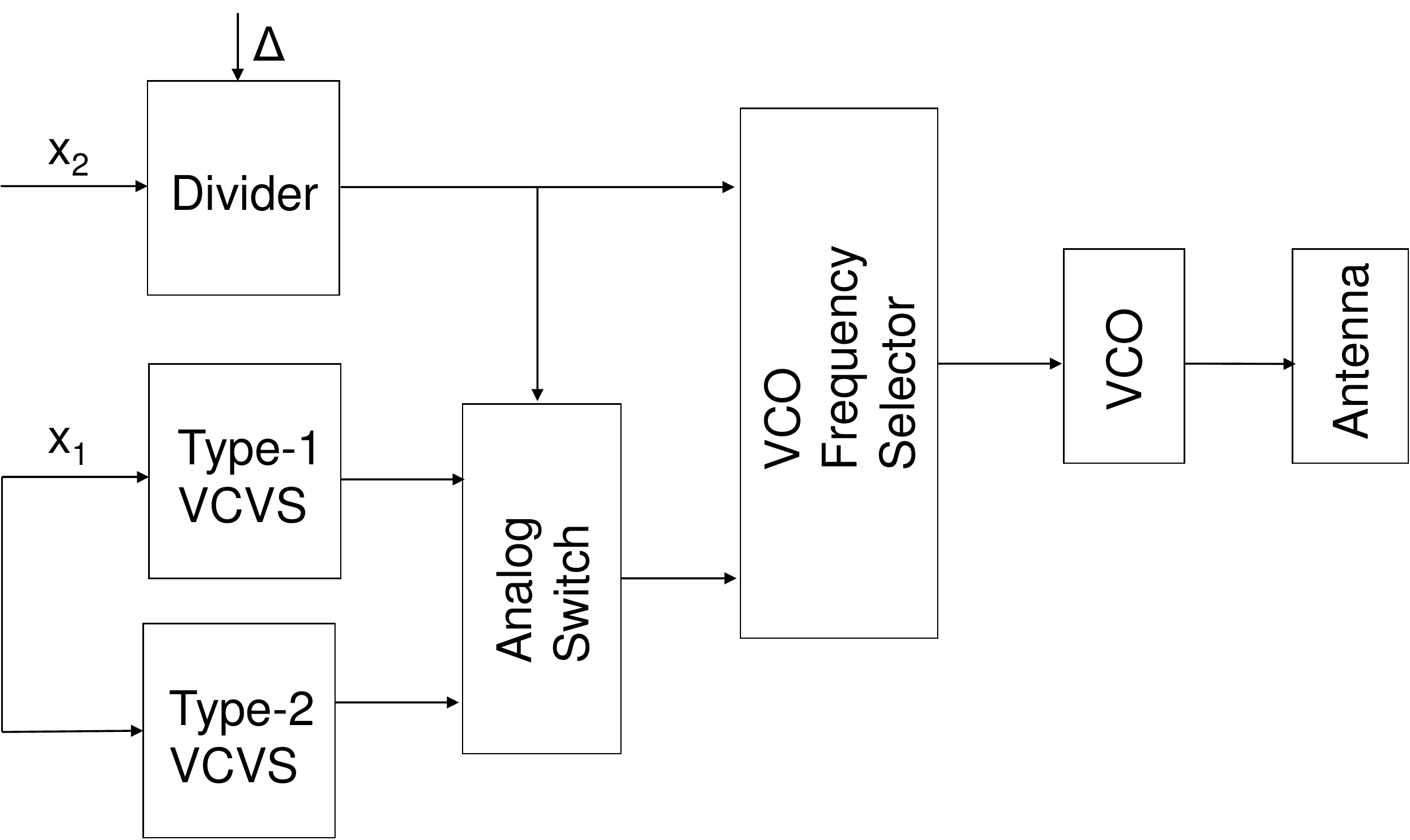}
        		\caption{}
        		\label{fig:Proposed_Transmitter}
        	\end{subfigure}%
~
        \begin{subfigure}[b]{0.25\textwidth}  
            \centering 
            \includegraphics[width=1\textwidth, height=3.5cm]{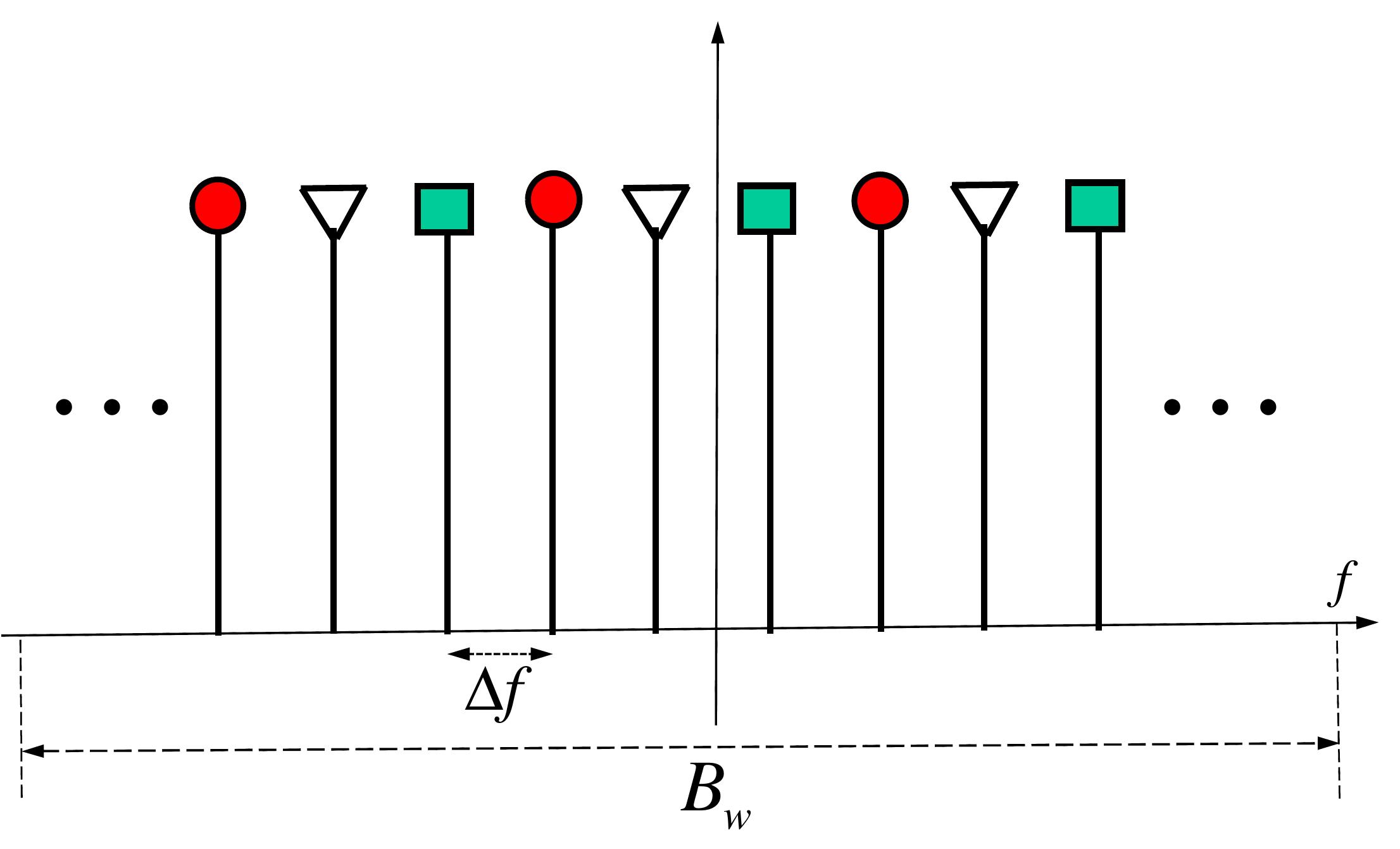}
            \caption{}
            \label{fig:freq_alloc}
        \end{subfigure}%
~
        \begin{subfigure}[b]{0.43\textwidth}   
            \centering 
            \includegraphics[width=1\textwidth, height=3.5cm]{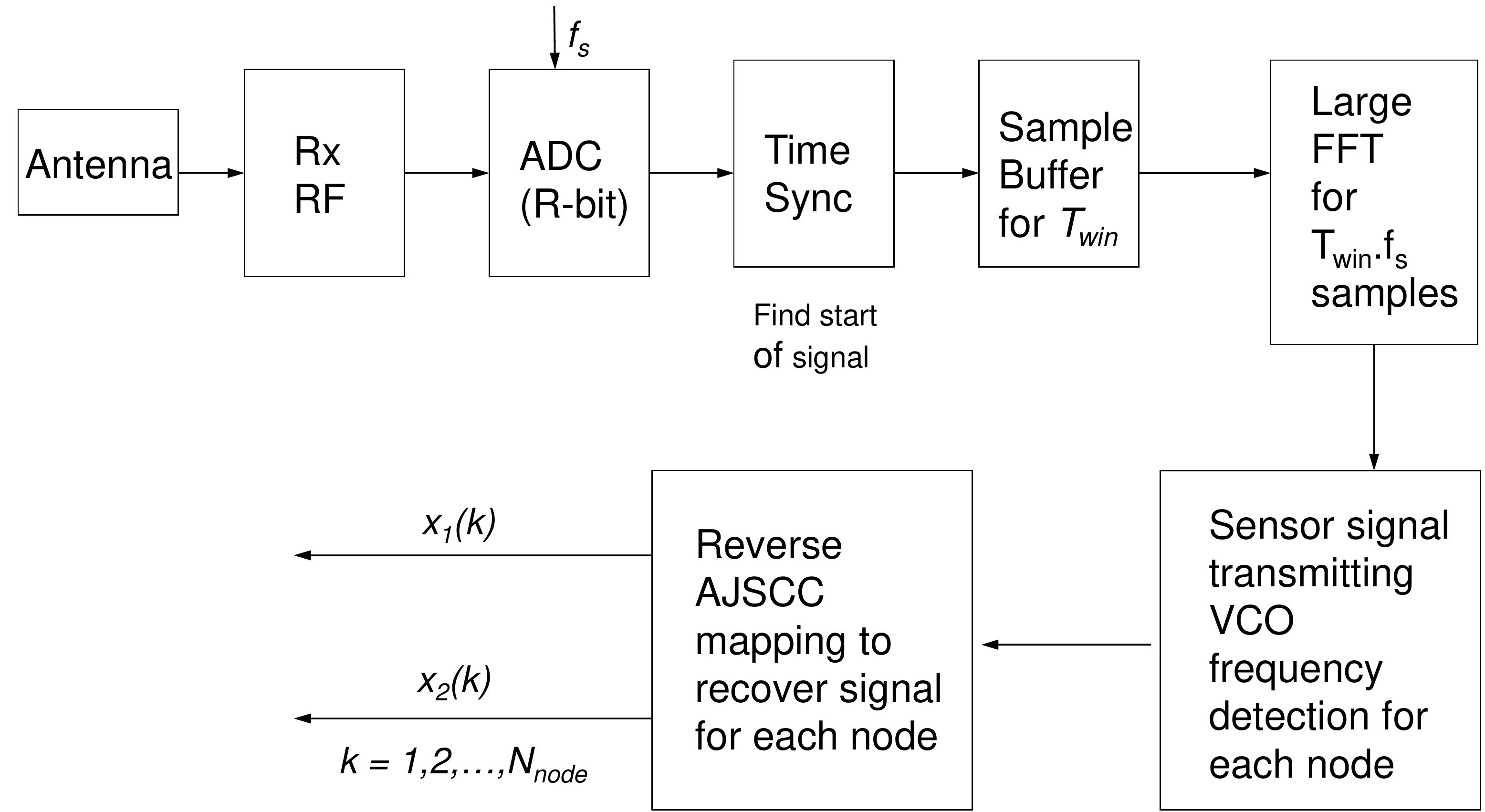}
            \caption{}
            \label{fig:Proposed_Receiver}
        \end{subfigure}
        \caption{\label{fig:app}(a) The proposed AJSCC device sensing two values $x_1,x_2$ is a synthetic mixed-signal circuit composed of analog signal compression (left-hand side) and nonlinear frequency position modulation (right-hand side). The outputs of analog signal compression control the frequency selection of VCO for generating RF signal; (b) Frequency Allocations in Frequency Position Modulation and Multiplexing~(FPMM); (c) The proposed FPMM receiver for receiving from a large number ($N_{node}$) of nodes at the same time.}
        \vspace{-0.15in}
\end{figure*}

\begin{table}[]
\small
\centering
\caption{Definitions of parameters used in the text.}
\label{table_parameters}
\begin{tabular}{m{0.5cm}m{7cm}}
\hline
$x_1$      & General notation of AJSCC input signal on x-axis                                                                                                          \\
$x_2$      & General notation of AJSCC input signal on y-axis                                                                                                          \\
$\Delta$      & General notation of spacing between AJSCC parallel lines on y-axis                                                                                                          \\
$V_T$      & Temperature Voltage of AJSCC input on x-axis, a specific type of signal of $x_1$                                                                           \\
$V_H$      & Humidity Voltage of AJSCC input signal on y-axis, a specific type of signal of $x_2$                                                                       \\
$V_R$      & Length of each horizontal line in voltage in AJSCC mapping for $V_T$ and $V_H$ compression                                                            \\
$\Delta_H$ & Spacing between parallel lines in AJSCC mapping for $V_T$ and $V_H$ compression                                                                           \\
$L$        & Number of parallel lines in AJSCC mapping                                      \\
$D_{max}$   & Maximum accumulated length of AJSCC mapping
\\
$B_w$      & Total double side baseband bandwidth of FPMM                                            \\
$f_s$      & Sampling frequency of ADC in the FPMM receiver                                     \\
$N_{node}$ & Total number of nodes in FPMM                                                       \\
$N_{0}$     & Number of quantization points for each parallel line
					\\
$N_{q}$    & Total number of quantization points in the final AJSCC mapped signal, $N_q = L N_0$ \\
$\Delta f$ & Frequency spacing of adjacent positions in FPMM, $\Delta f = B_w/(N_q N_{node} - 1) $                                      \\
$T_{win}$  & Time-domain window length in FPMM, $ T_{win} = 1/\Delta f $                           \\
$G$   & Receiver RF circuit power gain in $\mathrm{dB}$                                               \\
\hline
\end{tabular}
\vspace{-0.1in}
\end{table}

\subsection{Sensor Signal Modulation and Multiplexing}\label{sec:fpm}
In this section, we propose to use Frequency Position Modulation and Multiplexing~(FPMM) for sensors communicating to an aggregator node. The key design parameters for the sensor transmitter and FPMM receiver are summarized in Table~\ref{table_parameters}. Popular existing approaches to transmitting multiple signals from single-antenna sensors on the shared Radio Frequency~(RF) medium include Frequency Division Multiplexing~(FDM) or Time Division Multiplexing~(TDM) in which the signals are separated in distinct frequency bands or in time slots, respectively. However, the former allocates static frequency bands to sensors and as such is unfair to some of the sensors (interference-wise). Even though frequency hopping can be used to change the assigned static bands, synchronization is a problem especially in the analog domain. The latter is
hard to realize due to signal timing synchronization. Hence, we propose a novel multiplexing scheme~(FPMM) based on the interleaving of different user frequencies so as to be fair to all users. Below we first introduce FPMM (Sect.~\ref{sec:fpm1}) and explain its advantages along with how it can reduce the transmitter power (Sect.~\ref{sec:fpm_power}) while being robust to eavesdropping (Sect.~\ref{sec:fpm_scrambler}).

\subsubsection{Frequency Position Modulation and Multiplexing~(FPMM)}\label{sec:fpm1}
Consider we compress two continuous analog signals $x_1$ (e.g., blood temperature) and $x_2$ (e.g., blood pH) to one signal by AJSCC as in Fig.~\ref{fig:Proposed_Transmitter}. These signals are fed to the analog compression circuits composed of analog divider, two types of voltage-controlled voltage sources~(VCVS), and an analog switch. These circuits construct an analog joint source-channel coded~(AJSCC) circuit of the sensing signals $x_1$ and $x_2$. In the figure, parameter $\Delta$ is the line/stage spacing in AJSCC. More details on how these circuits perform AJSCC of the two signals are found in~\cite{Zhao16}. The AJSCC output controls the frequency of the voltage controlled oscillator~(VCO), which then modulates the signal to RF.

The basic principle of the proposed FPMM is depicted in Fig.~\ref{fig:freq_alloc}. Consider the FPMM has double-side bandwidth $B_w$ and the total number of multiplexed nodes/AJSCC devices is $N_{node}$. Let us denote the number of quantization levels of the final AJSCC-mapped signal as $N_q$. Hence, there will be $N_q N_{node}$ frequency positions, and the inter-position spacing is $\Delta f = B_w/(N_q N_{node} -1 )$. The AJSCC-mapped signal is represented by one pulse per node, at interleaved frequency locations without interfering to each other. In Fig.~\ref{fig:freq_alloc}, different colored arrows represent different nodes. In this example, three users are shown in an interleaving pattern with adjacent points spaced by $\Delta f$. Because of this interleaving (\textit{multiplexing}), we can see that the frequency positions of a single node are spread across the entire bandwidth, $B_w$, which ensures fairness to all nodes in terms of interference at specific frequencies. In this system it is assumed that the inter-position spacing is the same for all senor nodes. This is because it is assumed that all nodes have the same structure and are sensing the same analog source at different locations.

A simple approach of \textit{modulating} the AJSCC outputs to frequencies is to use a linear mapping where lower values are mapped to lower frequencies and vice-versa. However this approach is prone to risk of eavesdropping and interference in case data from multiple devices is spatially correlated. To solve these two problems, a scrambling-based modulation approach is discussed in Sect.~\ref{sec:fpm_scrambler}. If the receiver has a frequency-domain resolution of $\Delta f = 1/ T_{win}$, the number of frequency points in the whole bandwidth is,
\begin{equation}
N_f  = \left\lfloor {\frac{{B_w }}{{\Delta f}}} \right\rfloor,
\end{equation}and the maximum number of nodes supported is expressed by,
\begin{equation}\label{eq:maxusers}
N_{node,max}^{}  = \left\lfloor {\frac{{N_f }}{{N_q }}} \right\rfloor.
\end{equation}

The proposed FPMM receiver is depicted in Fig.~\ref{fig:Proposed_Receiver}. The received and down-converted signal is first sampled with sampling frequency $f_s$ and recording time $T_{win}$. The samples are sent for Fourier analysis to detect the frequency positions and to determine if a miss detection has occurred. The miss detection is said to have occurred if the received signal frequency position is not the transmitted frequency position. If the number of users is below the maximum limit, as shown in~\eqref{eq:maxusers}, the frequency domain samples can be recovered assuming high SNR. While the number of users approaches this limit, the frequency domain signal power will leak into target user, and the recovery will become more difficult. In Table~\ref{table_parameters}, the parameter $L$ controls $x_2$'s granularity, while parameter $N_q$ controls the output AJSCC signal granularity, which in turn determines $x_1$'s granularity. $N_q$ should be chosen in such as way as to meet the temperature granularity requirement. In the current system we consider additive Gaussian noise channel and do not assume Doppler. The reason is that the sensors are assumed to be deployed at fixed locations, and the Doppler is a very small value, which is much smaller than the frequency spacing $\Delta f$. In this work we are focusing on the static monitoring scenario so it is assumed that the sensors are static with no Doppler.

\begin{figure}
\begin{center}
\includegraphics[width=2.8in]{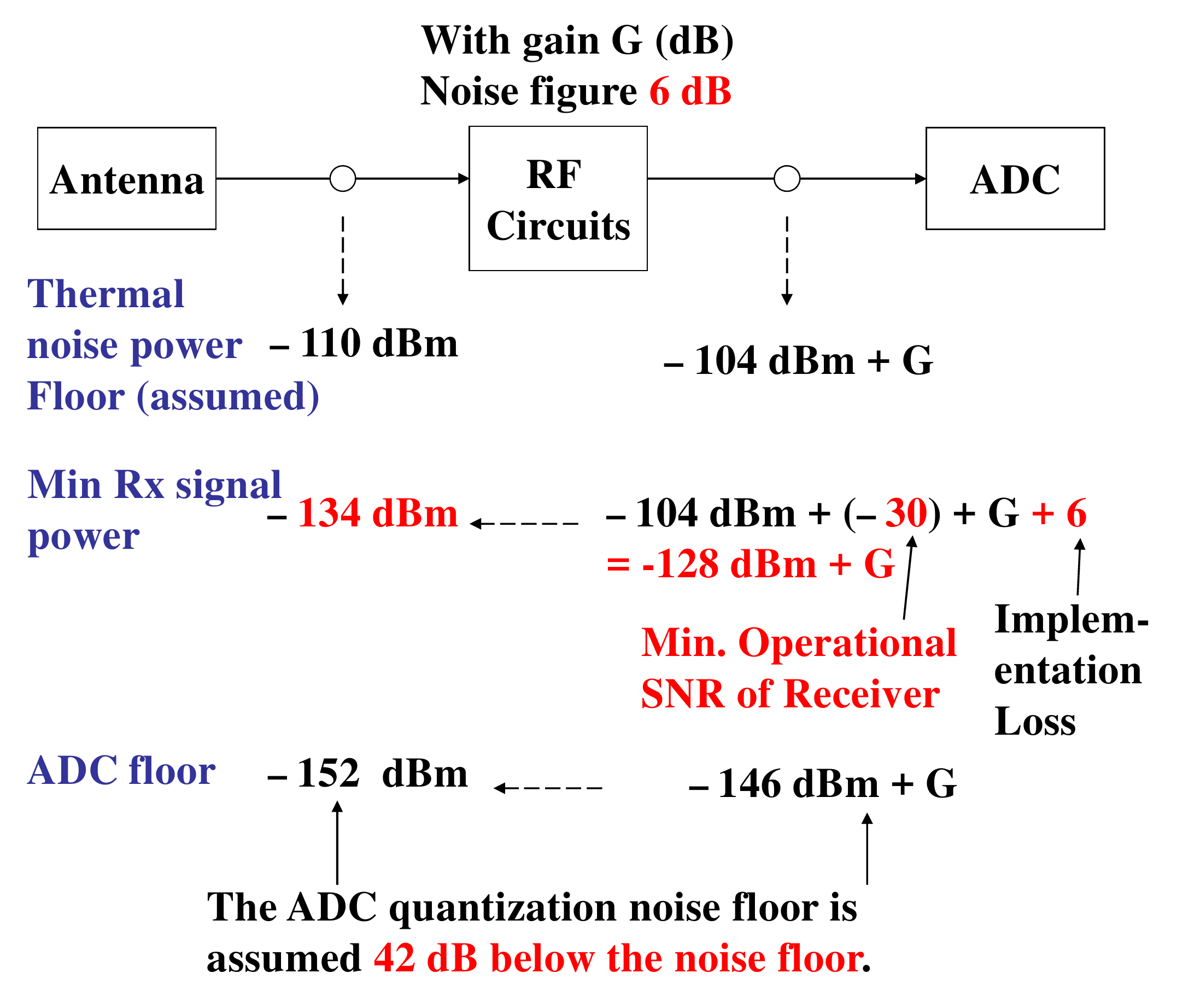}
\end{center}
\caption{Power analysis at FPMM receiver.}\label{fig:Rx_power_analysis}
\vspace{-0.2in}
\end{figure}

\begin{figure*}[t!]
        \centering   
           \begin{subfigure}[b]{0.33\textwidth}
        		\centering
        		\includegraphics[width=1\textwidth, height=2.5in]{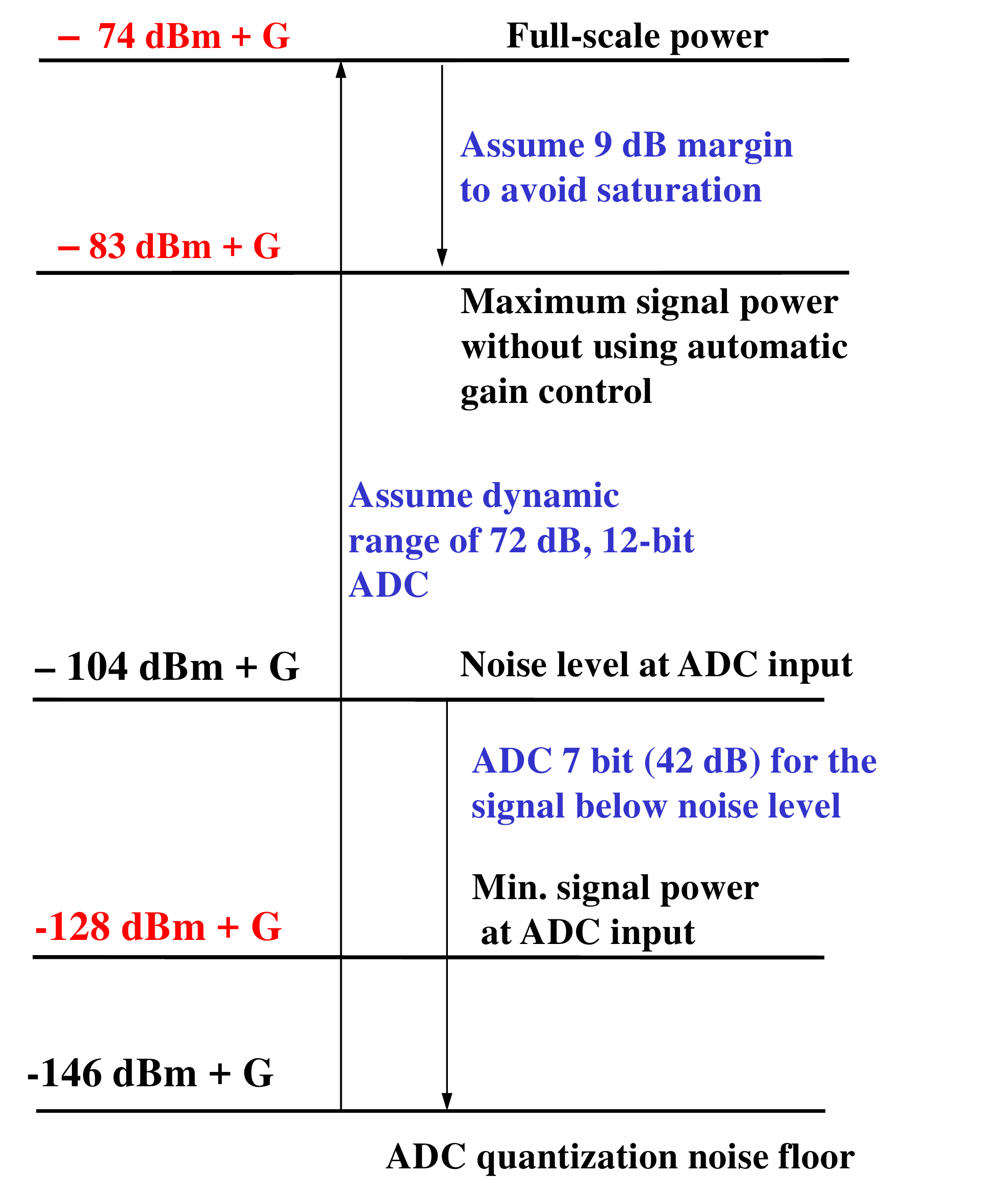}
        		\caption{}
        		\label{fig:DR_analysis}
        	\end{subfigure}%
~
        \begin{subfigure}[b]{0.33\textwidth}  
            \centering 
            \includegraphics[width=1\textwidth, height=2.3in]{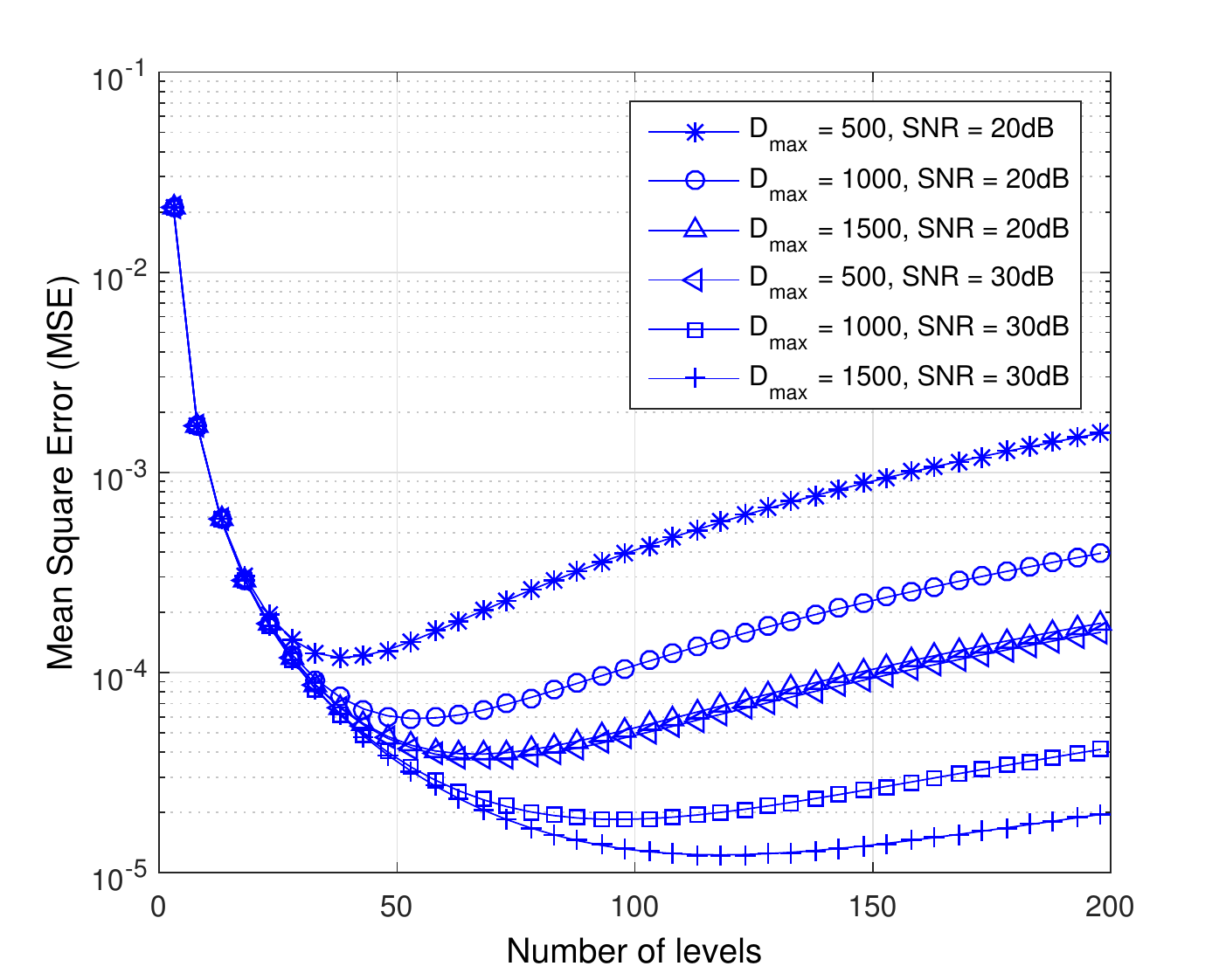}
            \caption{}
            \label{fig:MSE_vs_L_Dim_2}
        \end{subfigure}%
~
        \begin{subfigure}[b]{0.33\textwidth}   
            \centering 
            \includegraphics[width=1\textwidth,height=2.3in]{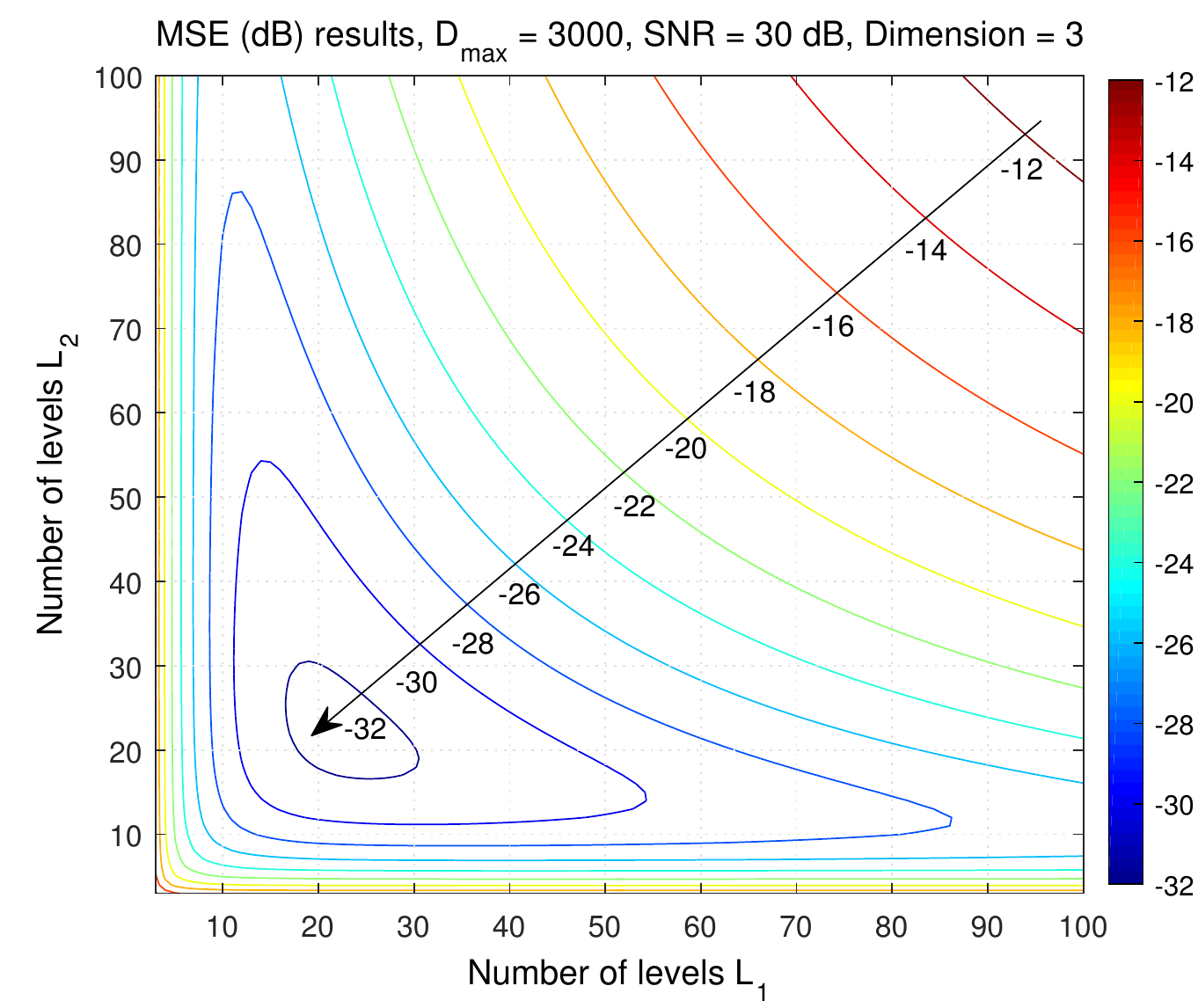}
            \caption{}
            \label{fig:MSE_vs_L_Dim_3}
        \end{subfigure}
        \caption{\label{fig:app}(a)~Dynamic range analysis at the ADC input; (b)~Mean Square Error~(MSE) vs. number of levels for analog signal compression by Shannon mapping (parameter $D_{max}$ is varied from $500$ to $1500$ for SNRs of $20$ and $30~\rm{dB}$ with dimension $N=2$); (c)~MSE contour graph vs. number of levels for 3-dimensional mapping ($N=3$), where $D_{max}=3000$ and SNR=$30~\rm{dB}$.}
     \vspace{-0.15in}
\end{figure*}

\subsubsection{Transmitter Power Reduction Using FPMM}\label{sec:fpm_power}
To analyze how the transmitting power can be reduced, we start with the signal power analysis at the FPMM receiver. The signal power values at the receiver antenna output and ADC input are depicted in Fig.~\ref{fig:Rx_power_analysis}. Assuming a thermal noise floor of -$110~\rm{dBm}$ at the antenna output, the thermal noise floor becomes -$104~\rm{dBm}$ + $G$, under the assumption of a RF circuit $G$ $\rm{dB}$ gain and a $6~\rm{dB}$ noise figure. We further assume the receiver minimum operational SNR is -$30~\rm{dB}$, and implementation loss of $6~\rm{dB}$. The minimum receiving signal power is thus -$128~\rm{dBm}$ + $G$ at the ADC input, or -$134~\rm{dBm}$ at antenna output. Assuming the ADC floor is $42~\rm{dB}$ below the thermal noise floor, then the ADC floor is -$146~\rm{dBm}$ + $G$ at the ADC input, or -$152~\rm{dBm}$ at the antenna output. Now, assume the receiver does not have Automatic Gain Control~(AGC). The received signal dynamic range at the ADC input is depicted in Fig.~\ref{fig:DR_analysis}. Given the ADC floor of -$146~\rm{dBm}$ + $G$ at the ADC input, and assuming a 12-bit ADC, which has a $72~\rm{dB}$ dynamic range, the full-scale power becomes -$74~\rm{dBm}$ + $G$. Furthermore, assume $9~\rm{dB}$ of margin for signal peak-to-average fluctuation and so as to avoid saturation, the maximum signal power at the receiver ADC input will be –$83~\rm{dBm}$ + $G$, again without the AGC. Meanwhile, the minimum signal power is also shown in the figure as -$128~\rm{dBm}$ + $G$.

From this power analysis it can be observed that the key factor to reduce the transmitting power is to reduce the minimum operational SNR at the receiver. It is found that such minimum operational SNR of our proposed system can be up to $30~\rm{dB}$ lower than in conventional digital linear and nonlinear modulation approaches, including FSK (in Z Wave), GFSK (in Bluetooth), and OQPSK (in ZigBee). The reason for our proposed system to achieve such a low SNR is that the transmitter and receiver adopted a structure that is completely different from the digital linear and nonlinear modulation approaches, which is explained as follows. First, we have the key observation that the sensing signal in the nature has a time period, during this period the signal can be treated as constant. For example, the temperature and humidity can be treated constant for $10$ seconds; or, if monitoring biomedical signals, e.g., the Glucose level, the sensing signal can be treated constant for $1$ second. However, current digital systems are sampling at much higher rates resulting in large redundancy as the readout signals are repeated for a large amount of samples. Digital modulation schemes in FSK, GFSK, and OQPSK are sampled with the receiver sampling rate to sense, process, and transmit the sensor outputs.

In our proposed system, the sensing signal is sensed in analog and continuously transmitted. The sampling of the sensing signal is done at the receiver side; and at the receiver, the samples are firstly buffered for the period in which the sensing signal can be treated as constant, then the buffered data is processed, as depicted in Fig.~\ref{fig:Proposed_Receiver},
to generate a one-time readout of the sensing source. This process has avoided the large redundancy compared to the digital sensing approaches, and more importantly allows for very low transmitting power. This one-time readout mechanism can collect all the signal power during the period of buffering to ensure signal detection. For example in our proposed receiver, large-sized FFT is adopted to perform frequency-domain analysis. The very small signal power can be recovered in the frequency domain with a large window/buffer time period. This has been verified by simulations. This buffer and one-time readout approach will allow for an ultra low operational SNR, around -$40$ to -$30~\rm{dB}$, which in turn will lead to an  ultra low transmitter power.

In Fig.~\ref{fig:Rx_power_analysis}, we have the minimum received signal power to be –$134~\rm{dBm}$ at the receiver antenna output. Assuming a path loss exponent of $3$, the power attenuation will be around $60~\rm{dB}$ for  a coverage of $100~\rm{m}$, and $90~\rm{dB}$ for $1~\rm{km}$. For the latter, the minimum transmitter signal power can be as low as -$44~\rm{dBm}$. Consider a medium power between –$134$ and -$89~\rm{dBm}$ (maximum signal power at antenna output). If we assume a –$112~\rm{dBm}$ medium received signal power, the medium transmitting power will be -$22~\rm{dBm}$. In contrast, the digital modulations such as FSK, GFSK, and OQPSK, all have an operational SNR around $0~\rm{dB}$, therefore the minimum received signal power will be –$104~\rm{dBm}$ at the receiver antenna output and minimum transmitter signal power for a $1~\rm{km}$ coverage is -$14~\rm{dBm}$ for these digital approaches. If we assume a –$82~\rm{dBm}$ medium received signal power, the medium transmitting power is $8~\rm{dBm}$ for these digital approaches for the $1~\rm{km}$ coverage. The verification of the minimum operational SNR of our proposal has been shown by simulation in Sect.~\ref{sec:perf_eval}.

\subsubsection{FPMM Scrambler to Mitigate Interference}\label{sec:fpm_scrambler}
Given a communication link with fixed frequency location interference, we can design a scrambler on the devices to change the frequency location of the modulated signal. Consider the application of collecting temperature and pulse oximeter readings from different parts of the body. This is needed for robustness and catching anomalies~\cite{Wu2011DataRecognition}. Given that the readings are collected from the same body, it is possible that some of them are correlated, generating similar AJSCC outputs and hence similar frequency-position modulated signals (frequencies) in the frequency domain. Now, if there is an interference just at the frequency location of the sensor signals, the signals of all those sensors will be distorted. 

To address this problem, a scrambler designed at the sensor nodes can re-allocate the frequency pulses to widely spaced locations. In this case, even if the measurement signals are very close, the frequency-position modulated signal will still be far apart in the frequency domain. The interference at a particular location can only affect the performance of a single sensor, while other sensors' signals will not be affected. The input of the scrambler is the AJSCC-modulated signal after quantization. The objective of the scrambler is to allocate the frequency location to another location that is statistically random and uniformly distributed in all the frequency positions. A pseudo-random number generator generating an integer from 1 to $N_q$ is assumed. The generated value is $x_r$. The AJSCC encoded and quantized signal is denoted as $x_q$; the following operation is then performed,
\begin{equation}
y=\mod(x_q + x_r, N_q).
\end{equation}
The frequency modulation is performed according to $y$ for the sensor. Different sensors have different seeds to generate pseudo-random numbers. The reverse operation can be performed at the receiver to de-scramble the signals, using the specific seed of each user. This approach is also robust to eavesdroppers due to the randomicity involved in mapping frequencies. Given that we are dealing with medical data in our application, this is an important step towards preserving user privacy. There is already some existing work on signal scrambling in the analog domain~\cite{seitner2008analog}.

\section{Performance Evaluation}\label{sec:perf_eval}
In this section, we evaluate the MSE performance of the generalized $N$-dimensional compression and the MDR performance of our proposed FPMM by varying SNR, bandwidth, and number of multiplexed devices under realistic conditions.

\textbf{N-dimensional Compression---MSE Analysis:}
\begin{figure*}[t!]
        \centering   
           \begin{subfigure}[b]{0.33\textwidth}
        		\centering
        		\includegraphics[width=1.1\textwidth]{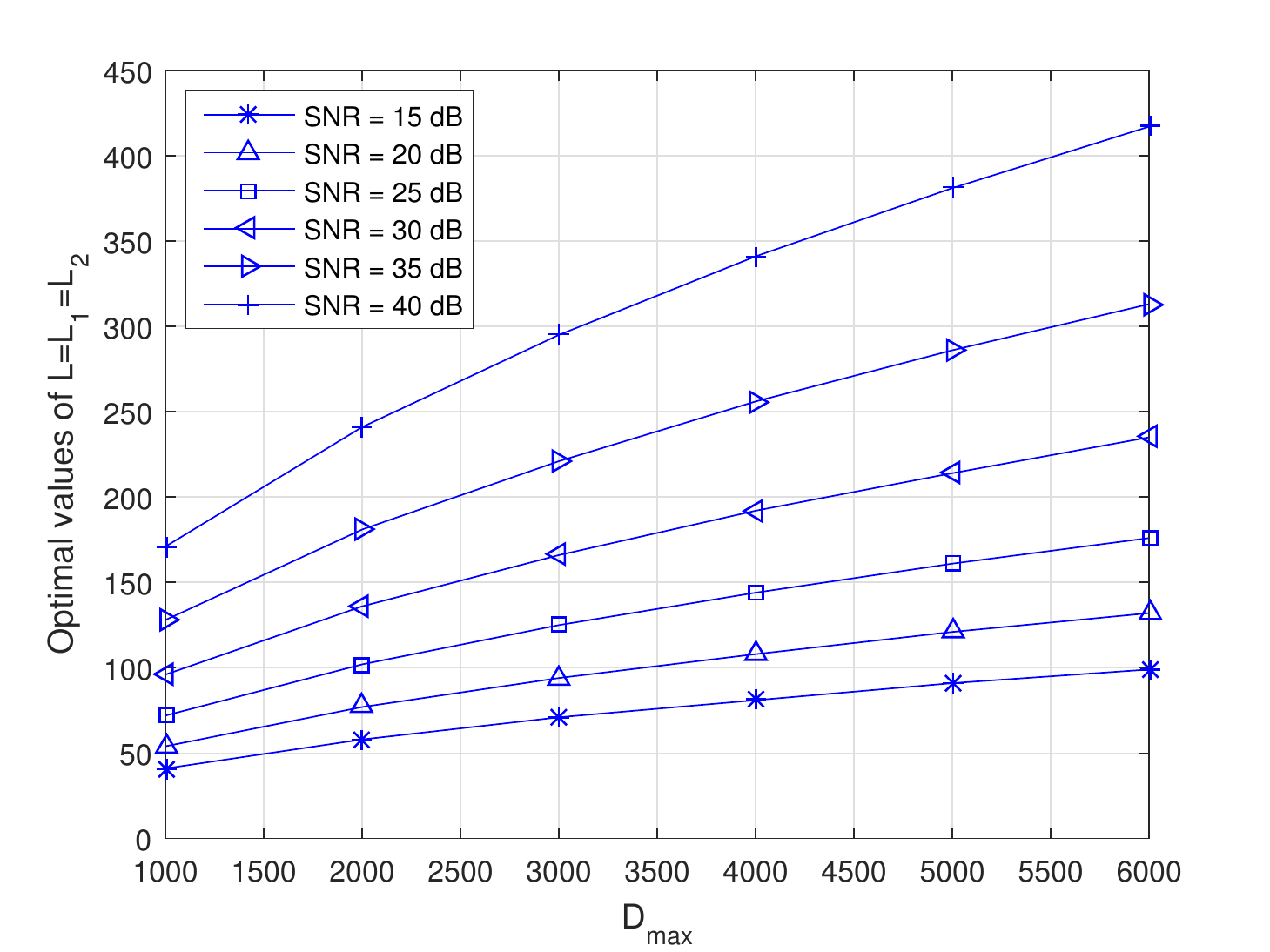}
        		\caption{}
        		\label{fig:Optimal_3D}
        	\end{subfigure}%
~
        \begin{subfigure}[b]{0.33\textwidth}  
            \centering 
            \includegraphics[width=1.1\textwidth]{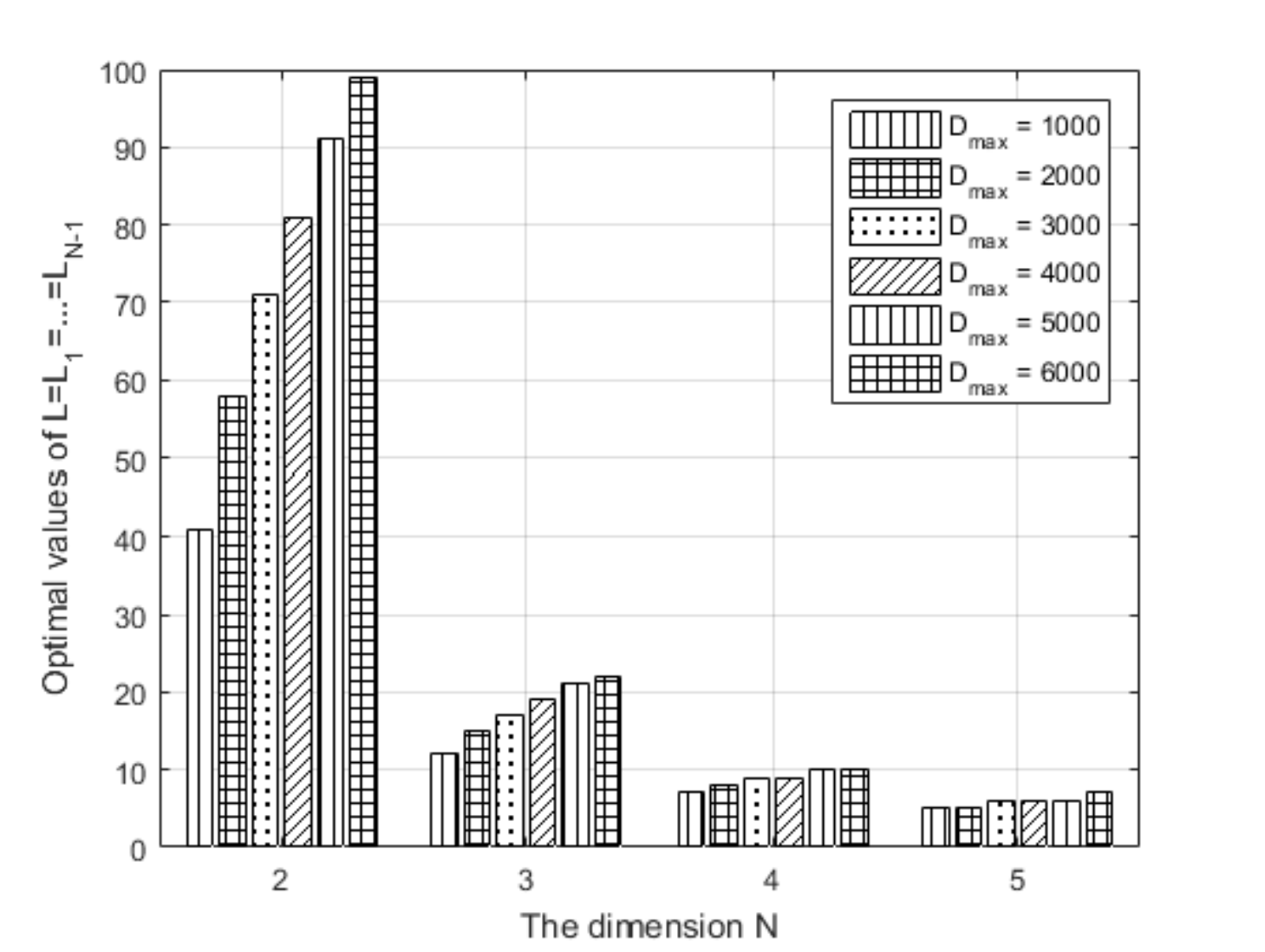}
            \caption{}
            \label{fig:L_vs_N_var_Dmax}
        \end{subfigure}%
~
        \begin{subfigure}[b]{0.33\textwidth}   
            \centering 
            \includegraphics[width=1.1\textwidth,height=1.95in]{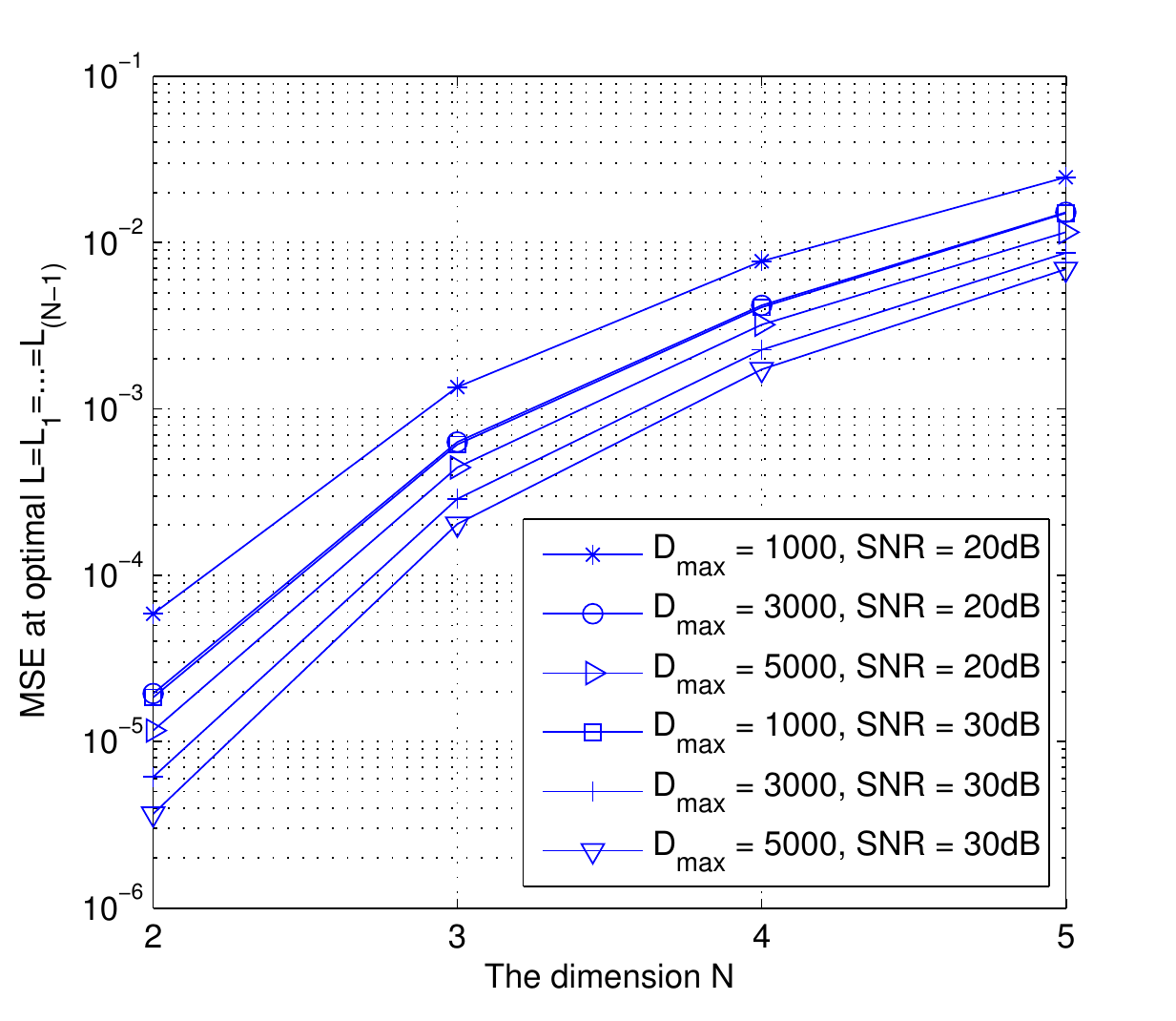}
            \caption{}
            \label{fig:MSE_vs_N_var_Dmax}
        \end{subfigure}
        \caption{\label{fig:app}(a)~Optimal $L_1$ and $L_2$ values, quantized to integers, of dimension $N=3$, for varying $D_{max}$ and SNR (note that, due to the symmetric structure of the problem, $L_1$ and $L_2$ have co-located optimal values); (b)~Optimal $L=L_1,...,L_{N-1}$ vs. number of dimensions $N$, for different $D_{max}$ values at SNR equal to $20~\rm{dB}$; (c)~Sum MSE at optimal $L=L_1,...,L_{N-1}$ vs. number of dimensions $N$, for different $D_{max}$ values for SNR equal to $20$ and $30~\rm{dB}$.}
        \vspace{-0.15in}
\end{figure*}
We studied the interesting tradeoff behavior between sum MSE and optimal number of stages for each dimension via MATLAB simulations. We know that there is a limit, $D_{max}$, on the mapped signal amplitude. Hence, we cannot arbitrarily increase the number of stages because as this number increases the range representing the first dimension reduces, leading to higher MSE for that dimension. In contrast, having a very small number of stages is also not desirable as such choice will introduce higher quantization error in the other quantized dimensions, leading to a higher sum MSE. We have studied this tradeoff behavior in simulations by varying the number of stages and $D_{max}$, and observed the resultant sum MSE. These results are shown in Fig.~\ref{fig:MSE_vs_L_Dim_2} for the 2-dimensional case (i.e., $N=2$, where one dimension is continuous while the other is discrete/quantized) and in Fig.~\ref{fig:MSE_vs_L_Dim_3} for the 3-dimensional case. From the results, as expected, we observe a local minimum for the MSE (due to the above-mentioned tradeoff), which gives the optimal number of stages for that particular $D_{max}$. This shows that, as $D_{max}$ decreases, the optimal number of lines/stages also decreases.

Figure~\ref{fig:MSE_vs_L_Dim_3} shows the MSE versus the number of levels for the 3-dimensional case (where one dimension is continuous and two are discrete/quantized) by varying the number of discrete levels in the second ($L_1$) and third ($L_2$) dimensions. We observe that there is a local minimum that gives the optimal number of levels, $L_{1,opt}=L_{2,opt} \approx 20$ for the second and third dimensions, for $D_{max}=3000$ at SNR=$30~\rm{dB}$. Note that these values change when $D_{max}$ and SNR change; yet a similar trend is observed
It is interesting to observe that the contour graph is symmetric by the two quantization levels, which indicates that $L_{1,opt}$ and $L_{2,opt}$ are co-located values.

Figure~\ref{fig:Optimal_3D} shows that (for the 3-dimensional case) these values increase, as $D_{max}$ or SNR increases. The reason for the optimal number of levels to change with $D_{max}$ and SNR is that these two parameters affect the quantization error and the Gaussian noise introduced, thus leading to errors at the receiver. With a large $D_{max}$, the error due to noise for the continuous signal $x_1$ is small, which allows for more parallel lines to be designed for the discrete signals $x_2$,...,$x_N$. For high SNR, the error due to Gaussian noise for the continuous signal $x_1$ is also small; thus, for a certain $D_{max}$, more lines are allowed for discrete signals $x_2$,...,$x_N$ to achieve the optimal MSE performance.

In Fig.~\ref{fig:L_vs_N_var_Dmax} we observe that, for a given $N$, optimal values of $L$ increase as $D_{max}$ increases, which is due to the reduced error on $x_1$; and the reason that the optimal number of levels decreases with increasing $N$, given a fixed $D_{max}$, is that the length for each dimension will dramatically reduce with increasing $N$. 
In Fig.~\ref{fig:MSE_vs_N_var_Dmax}, we can find that for large $D_{max}$ values, the MSE is around $10^{-4}\sim10^{-5}$ for $N=2$ and around $10^{-2}$ for $N=5$. The reason of this large MSE drop is because of the change in the number of quantization levels. The number of levels decreases by an order of magnitude going from dimension 2 to 5, as shown in
Fig.~\ref{fig:L_vs_N_var_Dmax}. We also note in Fig.~\ref{fig:MSE_vs_N_var_Dmax} that by increasing $D_{max}$ from 1000 to 5000, the MSE at dimension 4 and 5 are only moderately improved. These results indicate that there are intrinsic limitations imposed by the dimension that cap the MSE performance. By allowing higher dimensions (with quantization), there is an exponential demand on the maximum length $D_{max}$ as $N$ increases. On the other hand, as $D_{max}$ has the physical meaning of the modulated signal amplitude, it cannot be increased in such exponential manner due to power constraints. Such constraint on the maximum signal amplitude limits the total number of dimensions that our compression method can support.

\begin{figure*}[t!]
        \centering   
           \begin{subfigure}[b]{0.33\textwidth}
        		\centering
        		\includegraphics[width=1.1\textwidth]{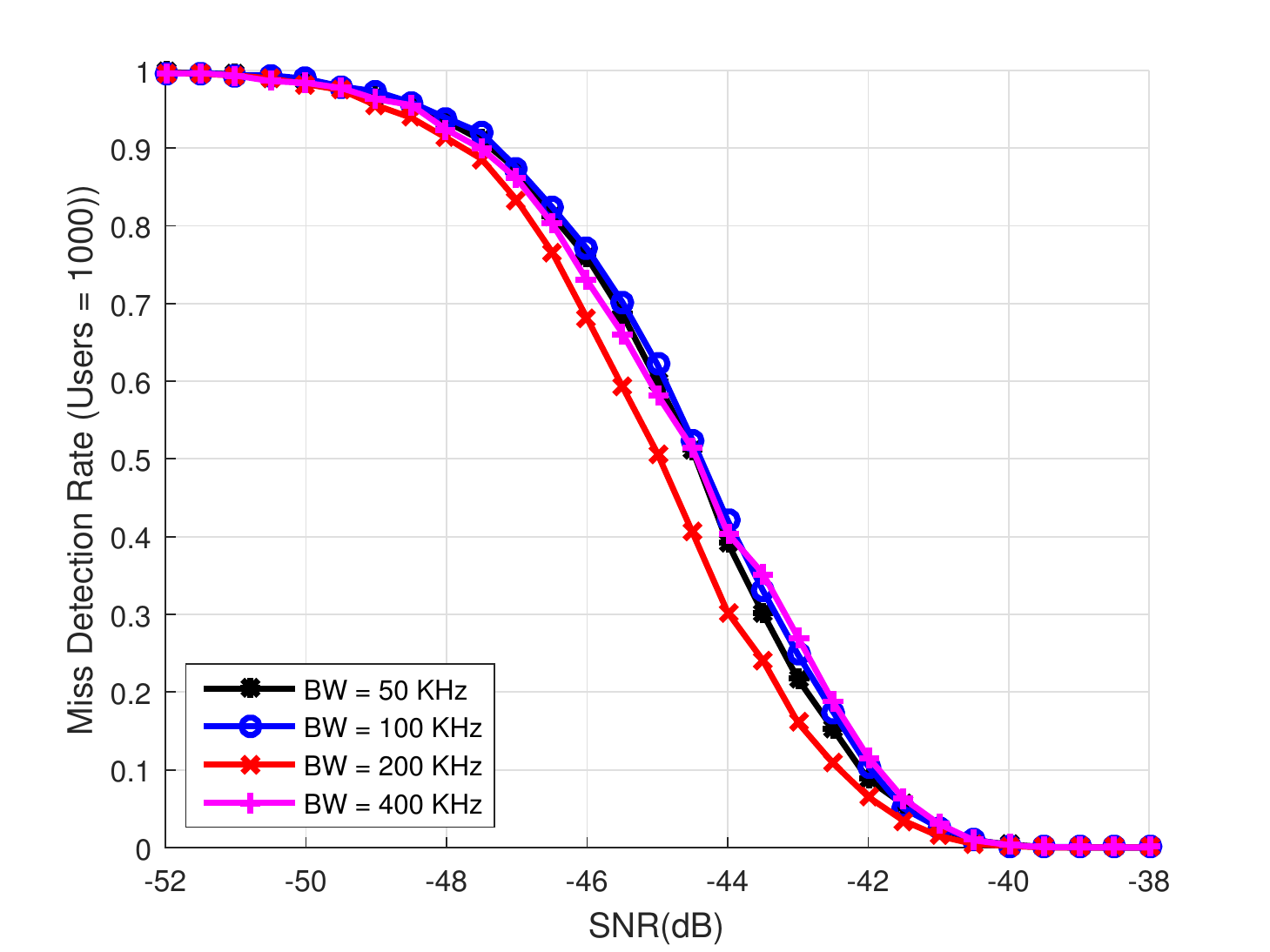}
        		\caption{}
        		\label{fig:mdr_var_snr}
        	\end{subfigure}%
~
        \begin{subfigure}[b]{0.33\textwidth}  
            \centering 
            \includegraphics[width=1.1\textwidth]{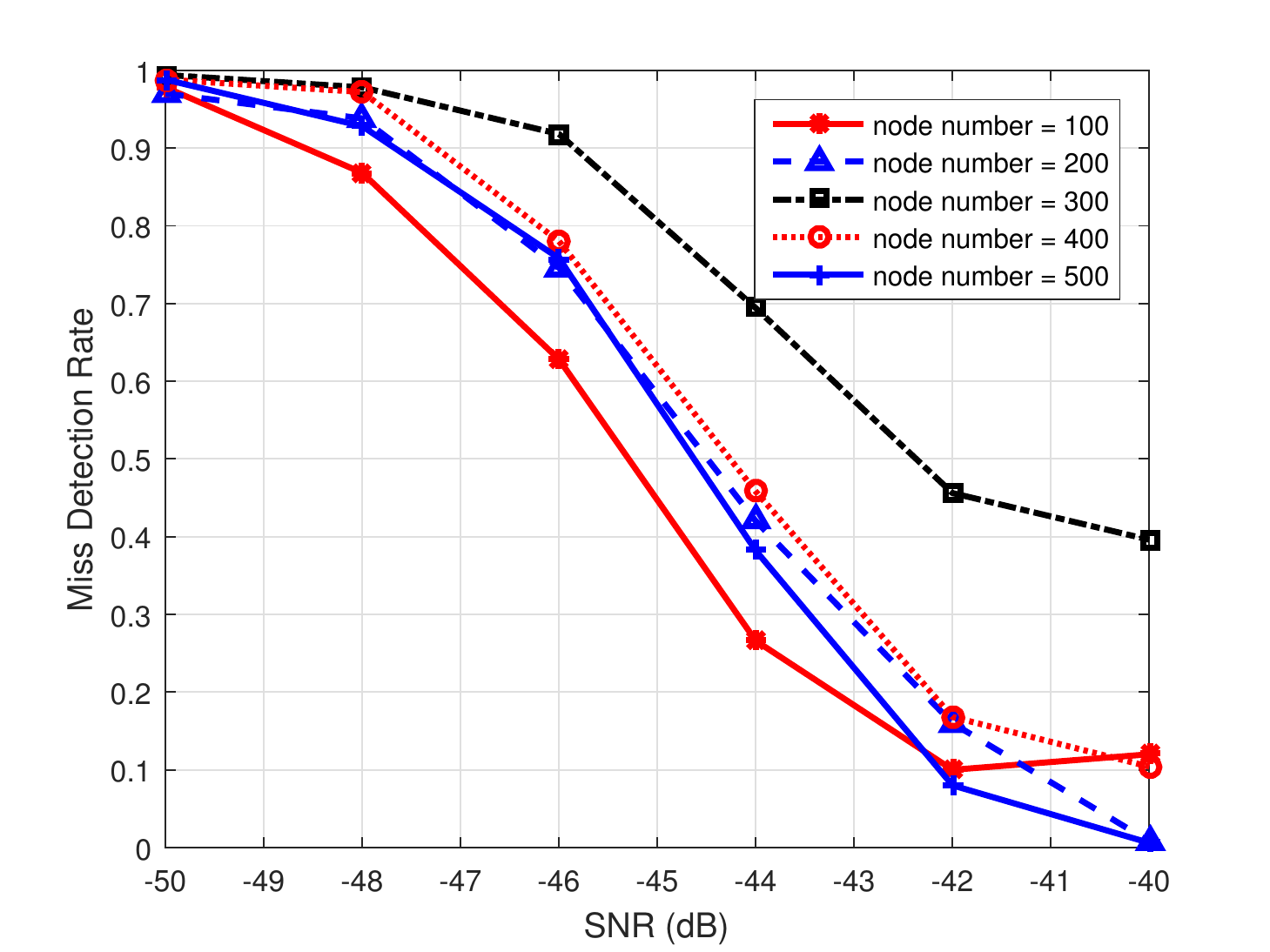}
            \caption{}
            \label{fig:awgn_50khz}
        \end{subfigure}%
~
        \begin{subfigure}[b]{0.33\textwidth}   
            \centering 
            \includegraphics[width=1.1\textwidth]{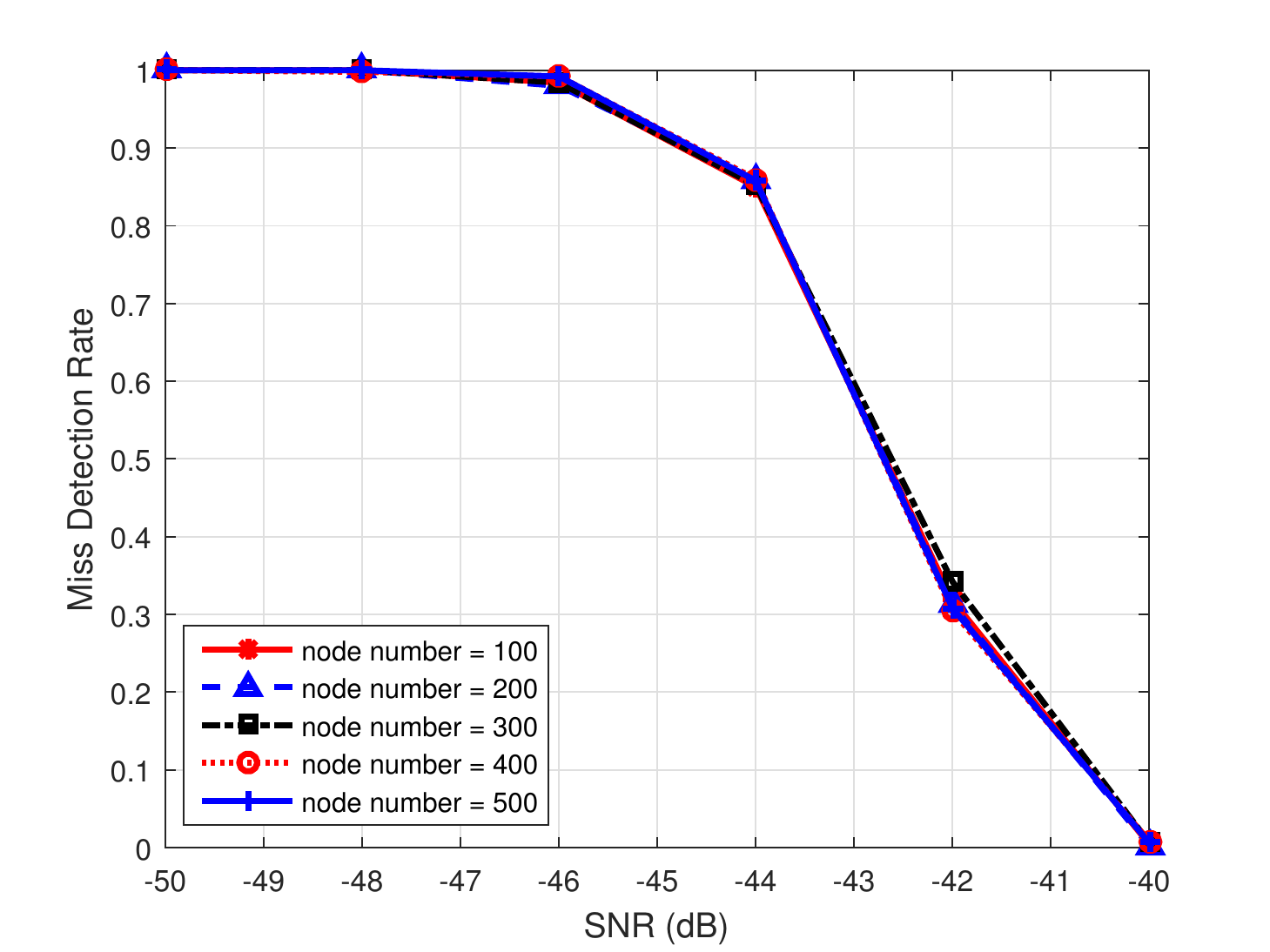}
            \caption{}
            \label{fig:awgn_1mhz}
        \end{subfigure}
        \caption{\label{fig:app}(a)~Miss Detection Rate~(MDR) vs. SNR for different BWs for the AWGN case (no. of users set to 1000); (b)~MDR vs. SNR for the AWGN case with different number of users (BW = $50~\rm{kHz}$); (c)~MDR vs. SNR for the AWGN case with different number of users (BW = $1~\rm{MHz}$).}
        \vspace{-0.15in}
\end{figure*}

\textbf{Frequency Position Modulation and Multiplexing---MDR Analysis:}
We studied the proposed FPMM-based multiplexing via MATLAB simulations.
We have carried out simulations to find the miss detection rate of the system for different bandwidth ($B_w$), number of users ($N_{user}$), and SNR. In practice, the collected samples are processed using a Fourier analysis system composed of cascaded Fourier transform blocks. Parameters $D_{max}$, $L$, and $N_q$ are decoupled from other parameters for detecting the frequency position modulated signal. When performing the simulations, parameters $D_{max}$, $L$, and $N_q$ are fixed. Our purpose is to find how the noise affects the detection, by varying bandwidth (sampling rate) and time of observation. These parameters are set to $L=50$, $D_{max}=5$, and $N_q=100$.

For recording the received signal, we define a time to record the FM modulated sequence as $T_{win}$, which is chosen to be $10~\rm{s}$. The received signal is firstly processed by FFT, then is detected based on the peak information to identify the frequency location of each node. Due to the analog nature of the transmitter, no pilot is available, therefore the standard digital detection techniques~\cite{Zhao08} are not applicable in this system. In these results, the metric of evaluation is the Miss Detection Rate~(MDR), defined as the number of users with detected frequency position not equaling the transmitting frequency position, divided by the total number of users multiplexed by FPMM. The SNR is defined as the signal power to noise power ratio at the receiver ADC input. We have set $N_{user}=1000$, and simulated the MDR by varying bandwidth from $50$ to $400~\rm{kHz}$ for the AWGN case. The results are shown in Fig.~\ref{fig:mdr_var_snr}. From the figure, we can notice that the MDR falls to a low value for SNR=-$40~\rm{dB}$ and is zero for SNR=-$30~\rm{dB}$. We also notice that the MDR behavior is approximately same for all bandwidths considered. For the AWGN case, due to the frequency modulation adopted, very low SNR can result in a detectable peak in frequency.

Considering AWGN channel with $50~\rm{kHz}$-bandwidth, the results of MDR vs. SNR for different number of users/nodes is shown in Fig.~\ref{fig:awgn_50khz}. The same metric for a $1~\rm{MHz}$-bandwidth is shown in Fig.~\ref{fig:awgn_1mhz}, whose purpose is to show how the MDR changes for a large bandwidth. We can see that the MDR is almost same for different number of nodes. This is because the number of nodes are under the limit of maximum allowed number of nodes and there is no interference by the proposed FPMM (since the bandwidth is large, the frequencies are far apart resulting in less interference and hence similar MDRs). We have also simulated the same results for a fading channel case. The results for $50~\rm{kHz}$ and $1~\rm{MHz}$ bandwidths are similar to the AWGN case. This is because FPMM, which is a non-linear modulation scheme, is intrinsically robust to fading. It can be observed from the results that the minimum operational SNR of the proposed system can be less than -$30~\rm{dB}$. Also, note that there is small difference between the numbers of users. This difference is negligible since the results show the same trend in the overlapped SNR region. 

\balance

\section{Conclusions and Future Work}\label{sec:conc}
We proposed multi-signal compression in the analog domain based on Analog Joint Source Channel Coding~(AJSCC). This addresses the problem of scalability of both data and number of sensors/things, in addition to power consumption. We also proposed a novel analog Frequency Position Modulation and Multiplexing~(FPMM) technique for multiple devices to communicate to an aggregator node (FPMM receiver) that is robust to interference and eavesdropping as well as being low power. The results showed the ability of these techniques to address the above mentioned challenges in the context of healthcare Internet of Things (IoT).

As future work, we will work on the following: (i)~AJSCC devices that are also able to receive configuration information from the aggregator; (ii)~generating configuration information such as quantization levels by leveraging machine learning at the aggregator; 
(iii)~analog scrambling; (iv)~increasing the spectral efficiency of our FPMM technique (all devices use the same bandwidth now) through spreading (code domain approach) and multiple receiver antennas (multi user MIMO) in interference environment~\cite{Zhao16a}; (v)~use different transformations such as wavelet, DWT, chirp, z-transform, DCT, and adopt different energy/power levels for different devices.

\bibliographystyle{IEEEtran}\small
\bibliography{reference_shannon,Mendeley,references_v3.0}

\end{document}